\documentclass[aps,showpacs,preprintnumbers,amsmath,amssymb,twocolumn,superscriptaddress,prl,floatfix,nofootinbib]{revtex4-2}

\usepackage[linesnumbered,lined,boxed,commentsnumbered,ruled,vlined]{algorithm2e}
\usepackage{algpseudocode}
\usepackage{amsfonts}
\usepackage{amsmath}
\usepackage{amssymb}
\usepackage{graphicx}
\usepackage{dcolumn}
\usepackage{bm,bbm}
\usepackage{xkcdcolors}
\usepackage{tabularx}
\usepackage{epstopdf}
\usepackage{xcolor}
\usepackage{mathrsfs}
\usepackage{subfigure}
\usepackage{caption}
\usepackage{mathtools}
\usepackage{float}
\usepackage{verbatim}
\usepackage{comment}
\usepackage{ragged2e}
\usepackage{physics}
\usepackage{hyperref}
\usepackage{url}
\usepackage{cleveref}
\usepackage{tikz}
\usepackage{utfsym}
\usepackage{multirow}
\usepackage{scalerel}

\DeclareCaptionJustification{justified}{\justifying}
\hypersetup{colorlinks=true, citecolor=orange, urlcolor=blue, linkcolor=magenta}

\definecolor{cel}{rgb}{0.0,0.53,0.74}
\definecolor{green}{rgb}{0.0,0.5,0.0}

\definecolor{colSDRG}{RGB}{51,117,56}
\definecolor{colSDCM}{RGB}{194,106,119}
\definecolor{colRec}{RGB}{230,158,0}

\definecolor{colpn}{RGB}{219,204,124}
\definecolor{aggbal}{RGB}{0,158,115}
\definecolor{aggunb}{RGB}{213,94,0}

\begin{document}

\title{Patterns of link reciprocity in directed, signed networks}

\author{Anna Gallo}
\email{anna.gallo@imtlucca.it}
\affiliation{IMT School for Advanced Studies, Piazza San Francesco 19, 55100 Lucca (Italy)}
\affiliation{INdAM-GNAMPA Istituto Nazionale di Alta Matematica `Francesco Severi', P.le Aldo Moro 5, 00185 Rome (Italy)}
\author{Fabio Saracco}
\affiliation{`Enrico Fermi' Research Center (CREF), Via Panisperna 89A, 00184 Rome (Italy)}
\affiliation{Institute for Applied Computing `Mauro Picone' (IAC), National Research Council, Via dei Taurini 19, 00185 Rome (Italy)}
\affiliation{IMT School for Advanced Studies, Piazza San Francesco 19, 55100 Lucca (Italy)}
\author{Renaud Lambiotte}
\affiliation{Mathematical Institute, University of Oxford, Woodstock Road, OX2 6GG Oxford (United Kingdom)}
\author{Diego Garlaschelli}
\affiliation{IMT School for Advanced Studies, Piazza San Francesco 19, 55100 Lucca (Italy)}
\affiliation{INdAM-GNAMPA Istituto Nazionale di Alta Matematica `Francesco Severi', P.le Aldo Moro 5, 00185 Rome (Italy)}
\affiliation{Lorentz Institute for Theoretical Physics, University of Leiden, Niels Bohrweg 2, 2333 CA Leiden (The Netherlands)}
\author{Tiziano Squartini}
\affiliation{IMT School for Advanced Studies, Piazza San Francesco 19, 55100 Lucca (Italy)}
\affiliation{INdAM-GNAMPA Istituto Nazionale di Alta Matematica `Francesco Severi', P.le Aldo Moro 5, 00185 Rome (Italy)}

\date{\today}

\begin{abstract}
Most of the analyses concerning signed networks have focused on the balance theory, hence identifying frustration with undirected, triadic motifs having an odd number of negative edges; much less attention has been paid to their directed counterparts. To fill this gap, we focus on signed, directed connections, with the aim of exploring the notion of frustration in such a context. When dealing with signed, directed edges, frustration is a multi-faceted concept, admitting different definitions at different scales: if we limit ourselves to consider cycles of length two, frustration is related to reciprocity, i.e. the tendency of edges to admit the presence of partners pointing in the opposite direction. As the reciprocity of signed networks is still poorly understood, we adopt a principled approach for its study, defining quantities and introducing models to consistently capture empirical patterns of the kind. In order to quantify the tendency of empirical networks to form either mutualistic or antagonistic cycles of length two, we extend the Exponential Random Graphs framework to binary, directed, signed networks with global and local constraints and, then, compare the empirical abundance of the aforementioned patterns with the one expected under each model. We find that the (directed extension of the) balance theory is not capable of providing a consistent explanation of the patterns characterising the directed, signed networks considered in this work. Although part of the ambiguities can be solved by adopting a coarser definition of balance, our results call for a different theory, accounting for the directionality of edges in a coherent manner. In any case, the evidence that the empirical, signed networks can be highly reciprocated leads us to recommend to explicitly account for the role played by bidirectional dyads in determining frustration at higher levels (e.g. the triadic one).
\end{abstract}


\maketitle

\section{I. Introduction}

\subsection{A. A brief history of balance}

The interest in the study of networks with positive, negative or missing edges can be traced back to the formulation of the so-called \emph{balance theory}, firstly proposed by Heider~\cite{heider1946attitudes} and further developed by Cartwright and Harary, who adopted \emph{signed graphs} to model it~\cite{cartwright1956structural}. The balance theory deals with the concept of balance: a complete, signed graph is said to be balanced if \emph{all triads} have an even number of negative edges, i.e. either zero (in this case, the three edges are all positive) or two. The so-called \emph{structure theorem} states that a complete, signed graph is balanced if and only if its set of nodes can be partitioned into $k=2$, disjoint subsets whose intra-modular edges are all positive and whose inter-modular edges are all negative. Cartwright and Harary extended the definition of balance to incomplete graphs~\cite{cartwright1956structural} by including cycles of length larger than three: a network is deemed as balanced if \emph{all cycles} have an even number of negative edges (although the points of each subset are no longer required to be connected). Taken together, the criteria above define the so-called (structural) \emph{strong balance theory}. Such a framework has been further extended by Davis~\cite{davis1967clustering}, who introduced the concept of \emph{$k$-balanced} networks, according to which signed graphs are balanced if their set of nodes can be partitioned into $k\geq2$, disjoint subsets with positive, intra-modular edges and negative, inter-modular edges. This generalised definition of balance has led to the formulation of the so-called (structural) \emph{weak balance theory}, according to which triads whose edges are all negative are balanced as well, since each node can be thought of as a group on its own. Taken together, the strong and the weak variants of the balance theory define the so-called (structural) \emph{traditional balance theory}: hence, $k$-balanced networks are traditionally balanced.\\

From a mesoscopic perspective, however, both versions of the balance theory require the presence of positive blocks along the main diagonal of the adjacency matrix ($k=2$, according to the strong variant; $k>2$, according to the weak variant) and of negative, off-diagonal blocks. As noticed in~\cite{doreian2009partitioning}, the block-structure defining the traditional balance theory is overly restrictive, dooming the vast majority of real-world, signed networks to be quickly dismissed as frustrated: in order to overcome what was perceived as a major limitation of the traditional balance theory, the proposal of replacing the (traditional) notion of frustration $F(\bm{\sigma})=L_\bullet^-+L_\circ^+$ - i.e. the total amount of misplaced connections, coinciding with the number of negative edges found within communities, $L_\bullet^-$, plus the number of positive edges found between communities, $L_\circ^+$ - with its softened variant $G(\bm{\sigma}|\alpha)=\alpha L_\bullet^-+(1-\alpha) L_\circ^+$ was advanced. Even ignoring the ambiguity due to the lack of a principled way for selecting $\alpha$ (the so-called `$\alpha$ problem' in~\cite{traag2019partitioning}), the criterion embodied by the $G$-test is still too strict~\cite{gallo2024assessing}. In the light of such a result, the second attempt pursued by Doreian and Mrvar to overcome the perceived limitations of the traditional balance theory was more radical, as they proposed to relax it by allowing for the presence of positive, off-diagonal blocks and negative, diagonal blocks - a generalisation that has gained the name of \emph{relaxed balance theory}~\cite{doreian2009partitioning}. Such a formulation of the relaxed balance theory lacks a proper mathematisation, as a score function such as $F(\bm{\sigma})$, or $G(\bm{\sigma}|\alpha)$, cannot be easily individuated. Besides, it is affected by the problem highlighted in~\cite{traag2009community}: `\emph{[\dots] if the number of clusters is left unspecified a priori, the best partition is the singletons partition (i.e. each node in its own cluster) [\dots]}'. A recent attempt of overcoming such a limitation is represented by the contribution~\cite{gallo2024assessing}: recasting the theory of balance within a statistical framework solves several problems at once, allowing an inference scheme for assessing if a signed graph is traditionally or relaxedly balanced to be defined.\\

\begin{figure}[t!]
\centering
\includegraphics[width=0.49\textwidth]{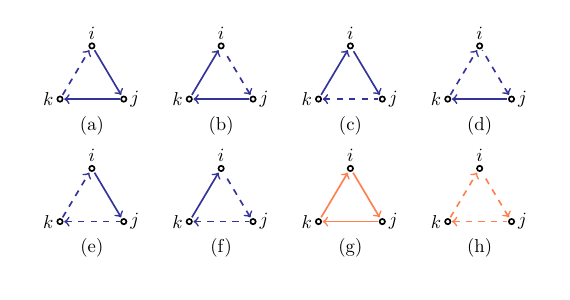}
\caption{Cyclic triads headed at node $i$: according to the status theory, blue triangles (a)$-$(f) are balanced while orange triangles (g) and (h) are unbalanced. While solid lines represent positive edges, dashed lines represent negative edges.}
\label{fig:1}
\end{figure}

The task of evaluating which variant of the traditional balance theory is best supported by empirical data has recently received a considerable attention: several metrics have been proposed to evaluate the level of balance~\cite{singh2017measuring,kirkley2019balance,estrada2014walk,estrada2019rethinking,kargaran2020quartic,siboni2022hybrid,talaga2023polarization}, or of the complementary notion of \textit{frustration}~\cite{aref2019balance,traag2019partitioning,aref2020multilevel,aref2020modeling}, and the question concerning the variant of the traditional balance theory that is best supported by data has been reformulated in statistical terms, by comparing the empirical abundance of the set of patterns that characterises each of them with the one predicted by a properly defined benchmark model~\cite{facchetti2011computing,saiz2017evidence,derr2018signed,huitsing2012univariate,lerner2016structural,becatti2019,fritz2022exponential,hao2024proper,gallo2024testing,gallo2024assessing}. Surprisingly, the answer crucially depends on a number of factors such as \emph{i)} the nature of the data, \emph{ii)} the measure adopted to quantify balance, \emph{iii)} the null model employed to carry out the analysis~\cite{gallo2024testing}: in particular, \emph{i)} null models induced by global constraints tend to favour the weak variant of the balance theory (according to which only the triangle with one, negative edge should be under-represented in real-world networks); \emph{ii)} null models induced by local constraints tend to favour the strong variant of the balance theory, according to which the triangle with three, negative edges should be under-represented as well. Biological networks, instead, seem to be characterised by a marked tendency towards frustration.

\begin{figure}[t!]
\centering
\includegraphics[width=0.49\textwidth]{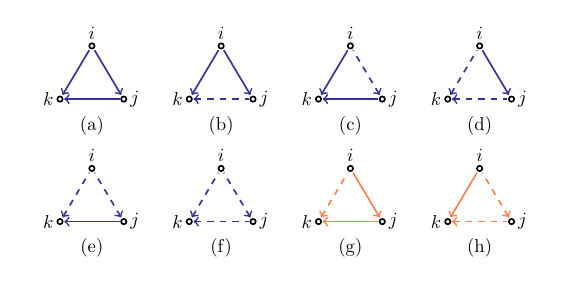}
\caption{Acyclic triads headed at node $i$: according to the status theory, blue triangles (a)$-$(f) are balanced while orange triangles (g) and (h) are unbalanced. While solid lines represent positive edges, dashed lines represent negative edges.}
\label{fig:2}
\end{figure}

\subsection{B. From \emph{balance} to \emph{status}}

Although popular, the balance theory disregards the direction of edges, a piece of information that is accounted for by the \emph{status theory}, implicitly proposed in~\cite{guha2004propagation}, where the meaning of the negative, directed edges was firstly discussed. There, the authors focused on the model called `web of trust', according to which users establish connections on the basis of trust/rating scores: `distrust' is, thus, modeled via a negative, directed edge, a methodological innovation allowing `distrust' to be explicitly distinguished from `absence of an opinion' - which, instead, is represented by the absence of a edge - and enhancing the prediction performance of the level of trust between pairs of users characterised by a missing rating. The status theory has been formally developed in~\cite{leskovec2010signed}: according to it, the sign of a edge between any two nodes depends on the (perceived) difference between their status. More precisely, the status theory affirms that a positive edge, directed from node $i$ to node $j$ indicates that the status of $j$ is perceived by $i$ as `higher' than its own status; conversely, a negative edge, directed from $i$ to $j$ indicates that the status of $j$ is perceived by $i$ as `lower' than its own status. Let us, now, consider a triangle with a directed, positive edge from $i$ to $j$, a directed, positive edge from $j$ to $k$ and a directed, negative edge from $k$ to $i$; such a triangle is unbalanced according to (both the strong and weak variant) of the traditional balance theory that, ignoring the directionality of the edges, depicts the usual `a friend of my friend is my enemy' but it is balanced according to the status theory: indeed, if $i$ believes its status to be `lower' than that of $j$ and $j$ believes its status to be `lower' than that of $k$, it is perfectly legitimate for $k$ to believe its status to be `higher' than that of $i$. Figures~\ref{fig:1} and~\ref{fig:2} depict the (directed, triadic) patterns classified as balanced (in blue) and unbalanced (in orange) according to the status theory.

The study of directed, signed networks has been approached from different perspectives and with different purposes: in~\cite{leskovec2010predicting,chiang2011exploiting,song2015link,ruiz2023triadic} the problems of edge and sign prediction are addressed; in~\cite{szell2010multirelational,aref2020multilevel,talaga2023polarization} a statistical validation of the level of frustration characterising directed, signed networks is carried out - to be noticed that the notion of frustration employed there is the one defined for undirected networks; in~\cite{girdhar2019social} the concept of \textit{status factor} is formalised, i.e. an index quantifying the social status of the nodes belonging to overlapping communities; in~\cite{linczuk2023multidimensional} the authors address the problem of opinion formation by labelling each edge with a sign that is determined by the similarity of the two, interacting agents; in~\cite{dinh2023enhancing} the level of frustration characterising directed, signed networks is proxied by the percentage of acyclic triads (not) satisfying the principle of transitivity - cyclic triads are, instead, excluded from the analysis.

\subsection{C. A brief history of reciprocity}

The study of reciprocity in binary, directed networks~\cite{holland1976local,wasserman1994social}, i.e. the tendency of vertex pairs to form mutual connections, has received an increasing attention during the years~\cite{garlaschelli2004patterns,garlaschelli2006multispecies,meyers2006predicting,boguna2005generalized,perra2009pagerank,zlatic2011model,garlaschelli2005structure,garlaschelli2010complex,zamora2008reciprocity,zlatic2009influence,stouffer2007evidence,squartini2011analytical}. Reciprocity has been shown to play a crucial role to classify~\cite{garlaschelli2004patterns} and model~\cite{garlaschelli2006multispecies} directed networks, understand the effects of a network structure on dynamical processes such as diffusion and percolation~\cite{meyers2006predicting,boguna2005generalized,perra2009pagerank}, explain patterns of growth in out-of-equilibrium networks such as Wikipedia~\cite{zlatic2011model} and the World Trade Web~\cite{garlaschelli2005structure,garlaschelli2010complex}, shape social relationships~\cite{evmenova2020analysis,evmenova2020structural}, study the onset of higher-order structures such as correlations~\cite{zamora2008reciprocity,zlatic2009influence} and triadic motifs~\cite{stouffer2007evidence,squartini2011analytical,milo2002network,squartini2012triadic}. Reciprocity also quantifies how much information is lost when a directed network is regarded as undirected: if the reciprocity of the original network is maximum, then the full information can be retrieved from the undirected projection; if, on the other hand, no reciprocity is observed, the uncertainty about the directionality of the original edges that have been converted into undirected ones is maximal~\cite{garlaschelli2004patterns}.\\

Generally speaking, directed networks range between two extremes, i.e. being purely bidirectional - such as the Internet, where information always travels both ways along computer cables - or being purely unidirectional - such as citation networks, where recent papers can cite less recent ones while the opposite cannot occur. A traditional way of quantifying where a real network lies is measuring the ratio of the number of edges pointing in both directions, $L^\leftrightarrow$, to the total number of edges, $L$,~\cite{wasserman1994social,newman2002email,serrano2003topology}, i.e. 

\begin{equation}\label{uns_rec}
r=\frac{L^\leftrightarrow}{L}=\frac{\sum_{i=1}^N\sum_{j(\neq i)}a_{ij}a_{ij}}{\sum_{i=1}^N\sum_{j(\neq i)}a_{ij}};
\end{equation}
clearly, $r=1$ for purely bidirectional networks while $r=0$ for purely unidirectional ones. Social networks~\cite{wasserman1994social}, email networks~\cite{newman2002email}, the WWW~\cite{newman2002email} and the World Trade Web~\cite{serrano2003topology} display an intermediate value of $r$.

In order to assess if mutual edges occur more or less (or just as) often than expected by chance, the empirical value of reciprocity must be compared with the one expected in a random graph with the same number of vertices~\cite{newman2002email}. To this aim, the definition of reciprocity reading

\begin{equation}
\rho=\frac{r-\langle r\rangle}{1-\langle r\rangle},
\end{equation}
with

\begin{equation}
\langle r\rangle=\frac{\langle L^\leftrightarrow\rangle}{\langle L\rangle}=\frac{\sum_{i=1}^N\sum_{j(\neq i)}\langle a_{ij}a_{ij}\rangle}{\sum_{i=1}^N\sum_{j(\neq i)}\langle a_{ij}\rangle}=\frac{N(N-1)p^2}{N(N-1)p}=p,
\end{equation}
was proposed. Still, such a measure suffers from two, major limitations: \emph{i)} it implements a comparison with the overall too simple benchmark known as Directed Random Graph Model~\cite{garlaschelli2004patterns}, according to which $p=L/N(N-1)$; \emph{ii)} does not admit a clear, statistical interpretation. In order to overcome them, researchers have replaced $\rho$ with the $z$-score

\begin{equation}
z[L^\leftrightarrow]=\frac{L^\leftrightarrow-\langle L^\leftrightarrow\rangle}{\sigma[L^\leftrightarrow]},
\end{equation}
where $\sigma[L^\leftrightarrow]=\sqrt{\text{Var}[L^\leftrightarrow]}$ and 

\begin{align}
\text{Var}[L^\leftrightarrow]&=\sum_{i=1}^N\sum_{j(>i)}\text{Var}[2a_{ij}a_{ji}]\nonumber\\
&=4\sum_{i=1}^N\sum_{j(>i)}p_{ij}p_{ji}(1-p_{ij}p_{ji}).
\end{align}

The $z$-score $z[L^\leftrightarrow]$ quantifies the number of standard deviations by which the empirical abundance of reciprocal edges differs from the expected one. After checking for the Gaussianity of $L^\leftrightarrow$ under the chosen benchmark - since it is a sum of dependent random variables, this is ensured by the generalisation of the Central Limit Theorem - a result $|z_m|\leq2$ ($|z_m|\leq3$) indicates that the empirical abundance of the pattern under study is compatible with the one expected under the chosen benchmark at the $5\%$ ($1\%$) level of statistical significance. On the other hand, a value $|z_m|>2$ ($|z_m|>3$) indicates that the empirical abundance of pattern $m$ is not compatible with the chosen benchmark at the same significance level: in the latter case, a value $z_m>0$ ($z_m<0$) indicates the tendency of the pattern to be over- (under-)represented in the data with respect to the null model. Real-world, directed networks have been shown~\cite{squartini2012triadic,squartini2013early} to display a value of reciprocity that neither the Directed Random Graph Model nor the Directed Configuration Model is capable of reproducing~\cite{squartini2012triadic,squartini2013early}.\\

The present paper is devoted to exploring the concept of signed reciprocity as well as its relationship with that of frustration. To this aim, we adopt a principled approach for its study, introducing models to quantify the tendency of empirical networks to form either mutualistic or antagonistic cycles of length two, that are suitable to analyse graphs with `plus one', `minus one' and `zero' edges as well as those with just `plus one' and `minus one' edges. More specifically, we extend the Exponential Random Graphs framework to binary, directed, signed networks with global and local constraints: as in the undirected case~\cite{gallo2024testing}, our approach has the clear advantage of being analytically tractable; besides, its versatility allows it to be employed either in presence of full information - to spot the degree of self-organisation of a signed network by detecting the patterns that are not explained by lower-level constraints~\cite{squartini2013early,Saracco2017,Becatti2019d,Caldarelli2020b,Neal2021} - or in presence of partial information - to infer the missing portion of a given signed network~\cite{Parisi2020}.

The rest of the paper is organized as follows. Section II introduces the formalism and the basic quantities we will consider in our analysis. Section III is devoted to the description of the maximum-entropy models we will employ to carry out our analysis. Section IV illustrates their application to a bunch of real-world networks. Section V discusses the main results of the paper and presents an outlook on future extensions of our work.

\section{II. Setting up the formalism}

In a signed graph, each edge can be positive, negative or missing. In what follows, we will focus on binary, directed, signed networks: each edge will be, thus, `plus one', `minus one' or `zero'. More formally, the generic entry of the adjacency matrix $\mathbf{A}$ reads $a_{ij}=-1,0,+1$. Since the total number of node pairs reads $N(N-1)$ and any `directed' node-pair can be `positively connected', `negatively connected' or `disconnected', the cardinality of the ensemble of binary, directed, signed graphs is $|\mathbb{A}|=3^{N(N-1)}$. Mathematical manipulations can be eased by employing Iverson's brackets to define the three functions of the entries reading $a_{ij}^-=[a_{ij}=-1]$, $a_{ij}^0=[a_{ij}=0]$ and $a_{ij}^+=[a_{ij}=+1]$: these new variables, inducing the definition of the matrices $\mathbf{A}^+$ and $\mathbf{A}^-$ (see Appendix~\hyperlink{AppA}{A}), are mutually exclusive, i.e. $\{a_{ij}^-,a_{ij}^0,a_{ij}^+\}=\{(1,0,0),(0,1,0),(0,0,1)\}$, and sum to 1, i.e. $a_{ij}^-+a_{ij}^0+a_{ij}^+=1$. The advantage of adopting such a formalism becomes evident when considering that each quantity is, now, computed on a matrix whose entries are, by definition, positive: as a consequence, it is positive as well.

\subsection{A. Total number of positive and negative links}

Our formalism allows us to define a number of quantities of interest: for instance, the number of positive and negative edges can be, respectively, defined as

\begin{align}
L^+=\sum_{i=1}^N\sum_{j(\neq i)}a_{ij}^+\quad\text{and}\quad L^-=\sum_{i=1}^N\sum_{j(\neq i)}a_{ij}^-;
\end{align}
naturally, $L=L^++L^-=\sum_{i=1}^N\sum_{j(\neq i)}(a_{ij}^++a_{ij}^-)$.

\subsection{B. Degree sequences}

The positive and the negative out-degree of node $i$ (i.e. the total number of positive and negative edges `outgoing from' node $i$) can be, respectively, defined as

\begin{align}
k_i^+=\sum_{\substack{j=1\\(j\neq i)}}^Na_{ij}^+\quad\text{and}\quad k_i^-=\sum_{\substack{j=1\\(j\neq i)}}^Na_{ij}^-
\end{align}
while the positive and the negative in-degree of node $i$ (i.e. the total number of positive and negative edges `entering into' node $i$) can be, respectively, defined as

\begin{align}
h_i^+=\sum_{\substack{j=1\\(j\neq i)}}^Na_{ji}^+\quad\text{and}\quad h_i^-=\sum_{\substack{j=1\\(j\neq i)}}^Na_{ji}^-;
\end{align}
naturally, $L^+=\sum_{i=1}^Nk_i^+=\sum_{i=1}^Nh_i^+$ and $L^-=\sum_{i=1}^Nk_i^-=\sum_{i=1}^Nh_i^-$.

\subsection{C. Reciprocity}

\begin{figure}[t!]
\centering
\includegraphics[width=0.49\textwidth]{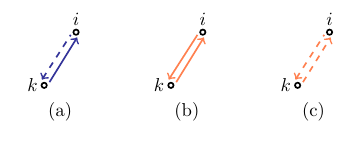}
\caption{Dyadic motifs for binary, directed, signed networks: (a) pair of edges, with discordant signs, pointing in opposite directions; (b) pair of positive edges pointing in opposite directions; (c) pair of negative edges pointing in opposite directions. According to the balance theory, pattern (a) is unbalanced while patterns (b) and (c) are balanced; according to the status theory, pattern (a) is balanced while patterns (b) and (c) are unbalanced.}
\label{fig:3}
\end{figure}

In the case of unsigned networks, the percentage of edges having a `companion' pointing in the opposite direction is defined as in Equation~\ref{uns_rec}; in the case of signed networks, four types of reciprocal patterns can be identified~\cite{shariff2013novel}, i.e. those leading to \emph{i)} \emph{positive reciprocity}, when nodes $i$ and $j$ trust each other; \emph{ii)} \emph{negative reciprocity}, when nodes $i$ and $j$ distrust each other; \emph{iii)} \emph{positive anti-reciprocity}, when node $i$ trusts node $j$ but node $j$ distrusts node $i$; \emph{iv)} \emph{negative anti-reciprocity}, when node $i$ distrusts node $j$ but node $j$ trusts node $i$. These patterns can be compactly represented via the family of dyadic motifs reading

\begin{align}
L^\leftrightarrow_+&=\sum_{i=1}^N\sum_{j(\neq i)}a_{ij}^+a_{ji}^+=2\sum_{i=1}^N\sum_{j(>i)}a_{ij}^+a_{ji}^+,
\end{align}
counting the pairs of positive edges pointing in opposite directions,

\begin{align}
L^\leftrightarrow_-&=\sum_{i=1}^N\sum_{j(\neq i)}a_{ij}^-a_{ji}^-=2\sum_{i=1}^N\sum_{j(>i)}a_{ij}^-a_{ji}^-,
\end{align}
counting the pairs of negative edges pointing in opposite directions and

\begin{align}
L^\leftrightarrow_\pm&=\sum_{i=1}^N\sum_{j(\neq i)}(a_{ij}^+a_{ji}^- + a_{ij}^-a_{ji}^+)\nonumber\\
&=2\sum_{i=1}^N\sum_{j(>i)}(a_{ij}^+a_{ji}^-+a_{ij}^-a_{ji}^+),
\end{align}
counting the pairs of edges, with different signs, pointing in opposite directions (see Figure~\ref{fig:3}). The tendency to establish anti-reciprocal (or single) connections can be quantified as well, upon defining

\begin{align}
L^\rightarrow_+&=\sum_{i=1}^N\sum_{j(\neq i)}a_{ij}^+(1-a_{ji}^+-a_{ji}^-)\nonumber\\
&=\sum_{i=1}^N\sum_{j(>i)}[a_{ij}^+(1-a_{ji}^+-a_{ji}^-)+a_{ji}^+(1-a_{ij}^+-a_{ji}^-)]
\end{align}
and

\begin{align}
L^\rightarrow_-&=\sum_{i=1}^N\sum_{j(\neq i)}a_{ij}^-(1-a_{ji}^+-a_{ji}^-)\nonumber\\
&=\sum_{i=1}^N\sum_{j(>i)}[a_{ij}^-(1-a_{ji}^+-a_{ji}^-)+a_{ji}^-(1-a_{ji}^+-a_{ij}^-)].
\end{align}

As a consequence, the following notions of reciprocity

\begin{align}
r^\leftrightarrow_+=\frac{L^\leftrightarrow_+}{L},\quad r^\leftrightarrow_-=\frac{L^\leftrightarrow_-}{L},\quad r^\leftrightarrow_\pm=\frac{L^\leftrightarrow_\pm}{L}
\end{align}
remain naturally defined, with $r^\leftrightarrow_+$ quantifying the percentage of edges appearing in reciprocal, positive pairs, $r^\leftrightarrow_-$ quantifying the percentage of edges appearing in reciprocal, negative pairs and $r^\leftrightarrow_\pm$ quantifying the percentage of edges appearing in a reciprocal pair with discordant signs; besides,

\begin{align}
r^\rightarrow_+=\frac{L^\rightarrow_+}{L},\quad r^\rightarrow_-=\frac{L^\rightarrow_-}{L}
\end{align}
with $r^\rightarrow_+$ quantifying the percentage of edges appearing in a positive, anti-reciprocal pair and $r^\rightarrow_-$ quantifying the percentage of edges appearing in a negative, anti-reciprocal pair. Similarly,

\begin{align}
s^\leftrightarrow_+=\frac{L^\leftrightarrow_+}{L^\leftrightarrow},\quad s^\leftrightarrow_-=\frac{L^\leftrightarrow_-}{L^\leftrightarrow},\quad s^\leftrightarrow_\pm=\frac{L^\leftrightarrow_\pm}{L^\leftrightarrow}
\end{align}
quantify the percentage of reciprocal edges that are both positive, the percentage of reciprocal edges that are both negative and the percentage of reciprocal edges with discordant signs.

As firstly noticed in~\cite{zajonc1965structural}, the concepts of reciprocity and frustration are not independent: in~\cite{talaga2023polarization}, the authors consider reciprocated dyads as generating two semi-cycles, whose degree of frustration is to be evaluated in the light of the balance theory; in~\cite{leskovec2010signed,evmenova2020analysis}, the authors notice that the balance theory has a larger statistical evidence on the subgraph defined by reciprocal edges. For the moment, we would simply like to stress that frustration and reciprocity are related by the meaning one attributes to (anti-)reciprocal patterns: if the collected data reflect a relationship such as `friendship', then panel (b) of Figure~\ref{fig:3} is balanced ($i$ sees $j$ as a friend, hence $a_{ij}=+1$, and $j$ sees $i$ as a friend, hence $a_{ji}=+1$); if, on the contrary, the collected data reflect a relationship like `subordination', then panel (b) of Figure~\ref{fig:3} is frustrated ($i$ sees $j$ as leading, hence $a_{ij}=+1$, but $j$ sees $i$ as leading as well, hence $a_{ji}=+1$).

\section{III. Benchmarks for binary, directed, signed networks}

Let us, now, generalise the Exponential Random Graphs (ERG) framework to binary, directed, signed networks. To this aim, we will follow the analytical approach introduced in~\cite{gallo2024testing} and carry out a constrained maximisation of Shannon entropy

\begin{align}
S=-\sum_{\mathbf{A}\in\mathbb{A}}P(\mathbf{A})\ln P(\mathbf{A})
\end{align}
where the sum runs over the ensemble of $|\mathbb{A}|=3^{N(N-1)}$ binary, directed, signed networks, a generic entry of which can assume the values $-1,0,+1$. Let us, now, discuss the models induced by two, different sets of constraints.

\subsection{A. Signed Directed Random Graph Model\\(SDRGM)}

The Signed Directed Random Graph Model (SDRGM) is induced by the Hamiltonian

\begin{align}
H(\mathbf{A})=\alpha L^+(\mathbf{A})+\beta L^-(\mathbf{A})
\end{align}
i.e. by the two, global constraints $L^+(\mathbf{A})$ and $L^-(\mathbf{A})$.

\subsubsection{1. Free-topology SDRGM}

According to the free topology variant of the SDRGM, each entry of a directed, signed network is a random variable whose behaviour is described by the finite scheme

\begin{align}
a_{ij}\sim
\begin{pmatrix}
-1 & 0 & +1\\
p^- & p^0 & p^+
\end{pmatrix},\quad\forall\:i\neq j
\end{align}
with

\begin{align}
p^-&\equiv\frac{e^{-\beta}}{1+e^{-\alpha}+e^{-\beta}}\equiv\frac{y}{1+x+y},\\
p^+&\equiv\frac{e^{-\alpha}}{1+e^{-\alpha}+e^{-\beta}}\equiv\frac{x}{1+x+y}
\end{align}
and $p^0\equiv1-p^--p^+$. In other words, $a_{ij}$ obeys a generalised Bernoulli distribution whose probability coefficients are determined by the (Lagrange multipliers of the) imposed constraints (see Appendix~\hyperlink{AppB}{B}): each positive edge appears with probability $p^+$, each negative edge appears with probability $p^-$ and each missing edge has a probability $p^0$.\\

In order to employ the SDRGM to study real-world networks, the parameters that define it need to be properly tuned: more specifically, one need to ensure that $\langle L^+\rangle_\text{SDRGM}=L^+(\mathbf{A}^*)$ and that $\langle L^-\rangle_\text{SDRGM}=L^-(\mathbf{A}^*)$. To this aim, the log-likelihood maximisation principle can be invoked~\cite{Garlaschelli2008}: it prescribes to maximise the function

\begin{align}
\mathcal{L}_\text{SDRGM}(x,y)&=\ln P_\text{SDRGM}(\mathbf{A}^*|x,y)\nonumber\\
&=\ln\left[\prod_{i=1}^N\prod_{j(\neq i)}(p^-)^{a_{ij}^-}(p^0)^{a_{ij}^0}(p^+)^{a_{ij}^+}\right]
\end{align}
with respect to the unknown parameters that define it. Such a recipe leads to the set of relationships

\begin{align}
p^+=\frac{L^+(\mathbf{A}^*)}{N(N-1)},\quad p^-=\frac{L^-(\mathbf{A}^*)}{N(N-1)}
\end{align}
and $p^0\equiv1-p^--p^+$.

\subsubsection{2. Fixed-topology SDRGM}

The way the SDRGM has been defined allows a network topological structure to vary along with the signs of the edges. A variant of the SDRGM that keeps the (directed) topology of the network under analysis fixed while (solely) randomising the signs of the edges is, however, definable. The role of random variables is, now, played by the entries of the adjacency matrix corresponding to connected pairs of nodes. Each of them obeys the finite scheme

\begin{align}
a_{ij}\sim
\begin{pmatrix}
-1 & +1\\
p^- & p^+
\end{pmatrix},\quad\forall\:i\neq j\:\:|\:\:|a_{ij}|=1
\end{align}
with

\begin{align}
p^-&\equiv\frac{e^{-\beta}}{e^{-\alpha}+e^{-\beta}}\equiv\frac{y}{x+y},\\
p^+&\equiv\frac{e^{-\alpha}}{e^{-\alpha}+e^{-\beta}}\equiv\frac{x}{x+y};
\end{align}
in other words, each entry satisfying $|a_{ij}|=1$ obeys a Bernoulli distribution whose probability coefficients are determined by the (Lagrange multipliers of the) imposed constraints (see Appendix~\hyperlink{AppB}{B}): each existing edge is assigned a `plus one' with probability $p^+$ and a `minus one' with probability $p^-$.\\

The maximisation of the log-likelihood function 

\begin{align}
\mathcal{L}_\text{SDRGM-FT}(x,y)&=\ln P_\text{SDRGM-FT}(\mathbf{A}^*|x,y)\nonumber\\
&=\ln\left[\prod_{i=1}^N\prod_{j(\neq i)}(p^-)^{a_{ij}^-}(p^+)^{a_{ij}^+}\right]
\end{align}
with respect to the unknown parameters that define it leads to the set of relationships

\begin{align}
p^+=\frac{L^+(\mathbf{A}^*)}{L(\mathbf{A}^*)},\quad p^-=\frac{L^-(\mathbf{A}^*)}{L(\mathbf{A}^*)}
\end{align}
with $L(\mathbf{A}^*)$ representing the (empirical) number of edges characterising the fixed topology under consideration.

\subsection{B. Signed Directed Configuration Model\\(SDCM)}

The two, aforementioned versions of the SDRGM are defined by global constraints; let us, now, consider a more refined null model, induced by local constraints. The SDCM is induced by the Hamiltonian

\begin{align}
H(\mathbf{A})=\sum_{i=1}^N[&\alpha_ik_i^+(\mathbf{A})+\beta_ik_i^-(\mathbf{A})+\gamma_ih_i^+(\mathbf{A})+\delta_ih_i^-(\mathbf{A})]
\end{align}
i.e. by the four vectors of local constraints $\{k_i^+(\mathbf{A})\}_{i=1}^N$, $\{k_i^-(\mathbf{A})\}_{i=1}^N$, $\{h_i^+(\mathbf{A})\}_{i=1}^N$ and $\{h_i^-(\mathbf{A})\}_{i=1}^N$.

\subsubsection{1. Free-topology SDCM}

According to the free topology variant of the SDCM, each entry of a directed, signed network is a random variable whose behaviour is described by the finite scheme

\begin{align}
a_{ij}\sim
\begin{pmatrix}
-1 & 0 & +1\\
p_{ij}^- & p_{ij}^0 & p_{ij}^+
\end{pmatrix},\quad\forall\:i\neq j
\end{align}
with

\begin{align}
p_{ij}^-\equiv\frac{e^{-(\beta_i+\delta_j)}}{1+e^{-(\alpha_i+\gamma_j)}+e^{-(\beta_i+\delta_j)}}\equiv\frac{y_iw_j}{1+x_iz_j+y_iw_j},\\
p_{ij}^+\equiv\frac{e^{-(\alpha_i+\gamma_j)}}{1+e^{-(\alpha_i+\gamma_j)}+e^{-(\beta_i+\delta_j)}}\equiv\frac{x_iz_j}{1+x_iz_j+y_iw_j}
\end{align}
and $p_{ij}^0\equiv1-p_{ij}^--p_{ij}^+$. In other words, $a_{ij}$ obeys a generalized Bernoulli distribution whose probability coefficients are determined by the (Lagrange multipliers of the) imposed constraints (see Appendix~\hyperlink{AppB}{B}): given any, two nodes $i$ and $j$, they are connected by a positive edge with probability $p_{ij}^+$, by a negative edge with probability $p_{ij}^-$ and are disconnected with probability $p_{ij}^0$.\\

In order to ensure that $\langle k_i^+\rangle_\text{SDCM}=k_i^+(\mathbf{A}^*)$, $\langle k_i^-\rangle_\text{SDCM}=k_i^-(\mathbf{A}^*)$, $\langle h_i^+\rangle_\text{SDCM}=h_i^+(\mathbf{A}^*)$, $\langle h_i^-\rangle_\text{SDCM}=h_i^-(\mathbf{A}^*)$, $\forall\:i$, let us maximise the log-likelihood function

\begin{align}
\mathcal{L}_\text{SDCM}(\mathbf{x},\mathbf{y},\mathbf{z},\mathbf{w})&=\ln P_\text{SDCM}(\mathbf{A}^*|\mathbf{x},\mathbf{y},\mathbf{z},\mathbf{w})\nonumber\\
&=\ln\left[\prod_{i=1}^N\prod_{j(\neq i)}(p_{ij}^-)^{a_{ij}^-}(p_{ij}^0)^{a_{ij}^0}(p_{ij}^+)^{a_{ij}^+}\right]
\end{align}
with respect to the unknown parameters that define it. Such a recipe leads to the set of relationships

\begin{align}
k_i^+(\mathbf{A}^*)&=\sum_{\substack{j=1\\(j\neq i)}}^N\frac{x_iz_j}{1+x_iz_j+y_iw_j}=\langle k_i^+\rangle,\quad\forall\:i,\\
k_i^-(\mathbf{A}^*)&=\sum_{\substack{j=1\\(j\neq i)}}^N\frac{y_iw_j}{1+x_iz_j+y_iw_j}=\langle k_i^-\rangle,\quad\forall\:i,\\
h_i^+(\mathbf{A}^*)&=\sum_{\substack{j=1\\(j\neq i)}}^N\frac{x_jz_i}{1+x_jz_i+y_jw_i}=\langle h_i^+\rangle,\quad\forall\:i,\\
h_i^-(\mathbf{A}^*)&=\sum_{\substack{j=1\\(j\neq i)}}^N\frac{y_jw_i}{1+x_jz_i+y_jw_i}=\langle h_i^-\rangle,\quad\forall\:i;
\end{align}
such a system can be solved only numerically (see Appendix~\hyperlink{AppC}{C} and Appendix~\hyperlink{AppD}{D}).

\subsubsection{2. Fixed-topology SDCM}

As for the SDRGM, a variant of the SDCM that keeps the (directed) topology of the network under analysis fixed while (solely) randomising the signs of the edges is definable. Again, the role of random variables is played by the entries of the adjacency matrix corresponding to connected pairs of nodes. Each of them obeys the finite scheme

\begin{align}
a_{ij}\sim
\begin{pmatrix}
-1 & +1\\
p_{ij}^- & p_{ij}^+
\end{pmatrix},\quad\forall\:i\neq j\:\:|\:\:|a_{ij}|=1
\end{align}
with

\begin{align}
p_{ij}^-&\equiv\frac{e^{-(\beta_i+\delta_j)}}{e^{-(\alpha_i+\gamma_j)}+e^{-(\beta_i+\delta_j)}}\equiv\frac{y_iw_j}{x_iz_j+y_iw_j},\\
p_{ij}^+&\equiv\frac{e^{-(\alpha_i+\gamma_j)}}{e^{-(\alpha_i+\gamma_j)}+e^{-(\beta_i+\delta_j)}}\equiv\frac{x_iz_j}{x_iz_j+y_iw_j};
\end{align}
in other words, each entry satisfying $|a_{ij}|=1$ obeys a Bernoulli distribution whose probability coefficients are determined by the (Lagrange multipliers of the) imposed constraints (see Appendix~\hyperlink{AppB}{B}): given any, two, connected nodes $i$ and $j$, their edge is assigned a `plus one' with probability $p_{ij}^+$ and a `minus one' with probability $p_{ij}^-$.\\

The maximisation of the log-likelihood function

\begin{align}
\mathcal{L}_\text{SDCM-FT}(\mathbf{x},\mathbf{y},\mathbf{z},\mathbf{w})&=\ln P_\text{SDCM-FT}(\mathbf{A}^*|\mathbf{x},\mathbf{y},\mathbf{z},\mathbf{w})\nonumber\\
&=\ln\left[\prod_{i=1}^N\prod_{j(\neq i)}(p_{ij}^-)^{a_{ij}^-}(p_{ij}^+)^{a_{ij}^+}\right]
\end{align}
with respect to the unknown parameters that define it leads to the set of relationships

\begin{align}
 k_i^+(\mathbf{A}^*)&=\sum_{\substack{j=1\\(j\neq i)}}^N|a^*_{ij}|\frac{x_iz_j}{x_iz_j+y_iw_j}=\langle k_i^+\rangle\quad\forall\:i,\\
 k_i^-(\mathbf{A}^*)&=\sum_{\substack{j=1\\(j\neq i)}}^N|a^*_{ij}|\frac{y_iw_j}{x_iz_j+y_iw_j}=\langle k_i^-\rangle\quad\forall\:i,\\
 h_i^+(\mathbf{A}^*)&=\sum_{\substack{j=1\\(j\neq i)}}^N|a^*_{ji}|\frac{x_jz_i}{x_jz_i+y_jw_i}=\langle h_i^+\rangle\quad\forall\:i,\\
 h_i^-(\mathbf{A}^*)&=\sum_{\substack{j=1\\(j\neq i)}}^N|a^*_{ji}|\frac{y_jw_i}{x_jz_i+y_jw_i}=\langle h_i^-\rangle\quad\forall\:i;
\end{align}
such a system can be solved only numerically (see Appendix~\hyperlink{AppC}{C} and Appendix~\hyperlink{AppD}{D}).

\subsection{C. The `SIMONA' Matlab package}

As an additional result, we release a Matlab-coded package that implements all the aforementioned models: its name is `SIMONA', an acronym standing for `Signed Models for Network Analysis', and is freely downloadable at this \href{https://it.mathworks.com/matlabcentral/fileexchange/167426-signed-models-for-network-analysis}{URL}.

\section{IV. Inference under binary, directed, signed benchmarks}

Let us, now, compare the empirical abundance of the quantities defined in the previous sections with the one expected under (any of) our null models. As anticipated, a useful indicator is the so-called $z$-score, i.e.

\begin{align}
z_m=\frac{N_m(\mathbf A^*)-\langle N_m\rangle}{\sigma[N_m]}
\end{align}
where $N_m(\mathbf A^*)$ is the empirical abundance of motif $m$ as measured on $\mathbf A^*$, $\langle N_m\rangle$ is its expected value under the chosen null model and $\sigma[N_m]=\sqrt{\langle N_m^2\rangle-\langle N_m\rangle^2}$ is the standard deviation of $N_m$ under the same null model. For some of the aforementioned quantities, the $z$-score can be explicitly computed. In particular, the expected values of the abundances of the reciprocal dyads read

\begin{align}
\langle L^\leftrightarrow_+\rangle&=2\sum_{i=1}^N\sum_{j(>i)}p_{ij}^+p_{ji}^+,\\
\langle L^\leftrightarrow_-\rangle&=2\sum_{i=1}^N\sum_{j(>i)}p_{ij}^-p_{ji}^-
\end{align}
and

\begin{align}
\langle L^\leftrightarrow_\pm\rangle&=2\sum_{i=1}^N\sum_{j(>i)}(p_{ij}^+p_{ji}^-+p_{ij}^-p_{ji}^+),
\end{align}
while the corresponding variances read

\begin{align}
\text{Var}[L^\leftrightarrow_+]&=4\sum_{i=1}^N\sum_{j(>i)}\text{Var}[a_{ij}^+a_{ji}^+]\nonumber\\
&=4\sum_{i=1}^N\sum_{j(>i)}p_{ij}^+p_{ji}^+(1-p_{ij}^+p_{ji}^+),\\
\text{Var}[L^\leftrightarrow_-]&=4\sum_{i=1}^N\sum_{j(>i)}\text{Var}[a_{ij}^-a_{ji}^-]\nonumber\\
&=4\sum_{i=1}^N\sum_{j(>i)}p_{ij}^-p_{ji}^-(1-p_{ij}^-p_{ji}^-)
\end{align}
and 

\begin{widetext}
\begin{align}
\text{Var}[L^\leftrightarrow_\pm]&=4\sum_{i=1}^N\sum_{j(>i)}\text{Var}[a_{ij}^+a_{ji}^-+a_{ij}^-a_{ji}^+]\nonumber\\
&=4\sum_{i=1}^N\sum_{j(>i)}\text{Var}[a_{ij}^+a_{ji}^-]+\text{Var}[a_{ij}^-a_{ji}^+]+2\text{Cov}[a_{ij}^+a_{ji}^-,a_{ij}^-a_{ji}^+]\nonumber\\
&=4\sum_{i=1}^N\sum_{j(>i)}[p_{ij}^+p_{ji}^-(1-p_{ij}^+p_{ji}^-)+p_{ij}^-p_{ji}^+(1-p_{ij}^-p_{ji}^+)-2p_{ij}^+p_{ji}^+p_{ij}^-p_{ji}^-]\nonumber\\
&=4\sum_{i=1}^N\sum_{j(>i)}[p_{ij}^+p_{ji}^-(1-p_{ij}^+p_{ji}^--p_{ij}^-p_{ji}^+)+p_{ij}^-p_{ji}^+(1-p_{ij}^-p_{ji}^+-p_{ij}^+p_{ji}^-)],
\end{align}
\end{widetext}
given that $\text{Cov}[a_{ij}^+a_{ji}^-,a_{ij}^-a_{ji}^+]=\langle a_{ij}^+a_{ji}^-a_{ij}^-a_{ji}^+\rangle-\langle a_{ij}^+a_{ji}^-\rangle\langle a_{ij}^-a_{ji}^+\rangle=-p_{ij}^+p_{ij}^-p_{ji}^+p_{ji}^-$, as the events $a_{ij}^+a_{ji}^-=1$ and $a_{ij}^-a_{ji}^+=1$ are mutually exclusive. The expected values and the variances of the abundances of the single dyads are provided in Appendix~\hyperlink{AppE}{E}. Let us remind that $z_m$ returns the number of standard deviations by which the empirical and the expected abundance of motif $m$ differ: while a result $|z_m|\leq2$ ($|z_m|\leq3$) would indicate that the empirical abundance of $m$ is compatible with the chosen null model at the $5\%$ ($1\%$) level of statistical significance, a result $|z_m|>2$ ($|z_m|>3$) would indicate that the empirical abundance of $m$ is not compatible with the chosen null model at the same significance level.

\begin{table*}[t!]
\centering
\begin{tabular}{c||c|c|c|c||c|c||c|c|c|c|c||c|c|c}
\hline
\textit{MMOG} & $N$ & $L$ & $L^+$ & $L^-$ & $c$ & $r$ & $r^\leftrightarrow_+$ & $r^\leftrightarrow_-$ & $r^\leftrightarrow_\pm$ & $r^\rightarrow_+$ & $r^\rightarrow_-$ & $s^\leftrightarrow_+$ & $s^\leftrightarrow_-$ & $s^\leftrightarrow_\pm$ \\
\hline
\hline
\textit{Day 10} & 1001 & 5448 & 5402 & 46 & $0.0054$ & 0.89 & 0.89 & 0.002 & 0.0011 & 0.10 & 0.006 & 0.99 & 0.003 & 0.0012 \\
\hline
\textit{Day 11} & 1079 & 6203 & 6118 & 85 & $0.0053$ & 0.88 & 0.87 & 0.004 & 0.0010 & 0.11 & 0.010 & 0.99 & 0.004 & 0.0011 \\
\hline
\textit{Day 12} & 1142 & 6833 & 6651 & 182 & $0.0052$ & 0.87 & 0.86 & 0.008 & 0.0015 & 0.11 & 0.018 & 0.99 & 0.009 & 0.0017 \\
\hline
\textit{Day 13} & 1220 & 7400 & 7152 & 248 & $0.0050$ & 0.87 & 0.86 & 0.012 & 0.0016 & 0.11 & 0.021 & 0.98 & 0.013 & 0.0019 \\
\hline
\textit{Day 14} & 1296 & 7936 & 7617 & 319 & $0.0047$ & 0.86 & 0.84 & 0.012 & 0.0025 & 0.11 & 0.027 & 0.98 & 0.014 & 0.0029 \\
\hline
\textit{Day 15} & 1328 & 8416 & 8055 & 361 & $0.0048$ & 0.86 & 0.85 & 0.014 & 0.0024 & 0.11 & 0.028 & 0.98 & 0.016 & 0.0028 \\
\hline
\textit{Day 16} & 1371 & 8789 & 8352 & 437 & $0.0047$ & 0.86 & 0.84 & 0.015 & 0.0036 & 0.11 & 0.033 & 0.98 & 0.017 & 0.0042 \\
\hline
\textit{Day 17} & 1414 & 9184 & 8646 & 538 & $0.0046$ & 0.85 & 0.83 & 0.016 & 0.0046 & 0.11 & 0.040 & 0.98 & 0.019 & 0.0054 \\
\hline
\textit{Day 18} & 1448 & 9451 & 8847 & 604 & $0.0045$ & 0.85 & 0.83 & 0.017 & 0.0042 & 0.11 & 0.045 & 0.98 & 0.020 & 0.0050 \\
\hline
\textit{Day 19} & 1489 & 9780 & 9104 & 676 & $0.0044$ & 0.84 & 0.82 & 0.018 & 0.0037 & 0.11 & 0.049 & 0.97 & 0.021 & 0.0044 \\
\hline
\textit{Day 20} & 1520 & 10109 & 9384 & 725 & $0.0044$ & 0.84 & 0.82 & 0.018 & 0.0042 & 0.11 & 0.052 & 0.97 & 0.021 & 0.0049 \\
\hline
\end{tabular}
\caption{Descriptive statistics of the largest connected component of each of the eleven snapshots of the MMOG dataset~\cite{szell2010multirelational}. The table shows the total number of nodes, $N$, the total number of edges, $L$, the total number of positive and negative edges, $L^+$ and $L^-$, the edge density, $c=L/N(N-1)$, the percentage of reciprocated edges, $r=L^{\leftrightarrow}/L$, and the empirical value of the reciprocity measures that we have introduced. Overall, these networks are very sparse but very reciprocated: the few, existing edges are paired in dyads the vast majority of which has a $+/+$ signature.}
\label{tab:1}
\end{table*}

\begin{figure*}[t!]
\centering
\includegraphics[width=\textwidth]{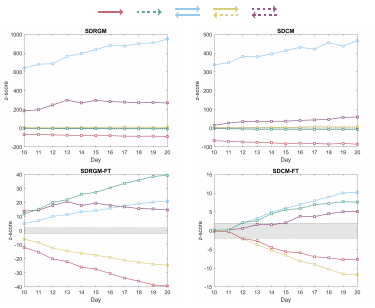}
\caption{Top panels: evolution of the $z$-scores of our dyadic motifs, under free-topology benchmarks (the SDRGM, on the left, and the SDCM, on the right), for the MMOG dataset. The $z$-scores induced by the SDRGM are larger, in absolute value, than the $z$-scores induced by the SDCM, an evidence indicating that accounting for the heterogeneity of nodes greatly enhances the explanatory power of a model. Moreover, the reciprocal dyads with concordant signs (in light-blue and purple) are over-represented in the data while the single, positive dyads (in red) are under-represented in the data. Bottom panels: evolution of the $z$-scores of our dyadic motifs, under fixed-topology benchmarks (i.e. the SDRGM-FT and the SDCM-FT), for the MMOG dataset. While the explanatory power of the fixed-topology benchmarks is larger than that of their free-topology counterparts, the reciprocal dyads with concordant signs (in light-blue and purple) and the single, negative dyads (in green) are over-represented in the data while the reciprocal dyads with discordant signs (in yellow) and the single, positive dyads (in red) are under-represented in the data. While the free-topology benchmarks favour the directed version of the balance theory, the fixed-topology benchmarks lead to contradictory results. The gray area corresponds to $z\in[-2,+2]$.}
\label{fig:4}
\end{figure*}

\section{V. Datasets description}

Let us, now, provide a brief description of the three datasets we have considered for the present analysis.\\

\paragraph{MMOG dataset.} This dataset collects information about the relationships among the $\simeq300.000$ players of a massive multiplayer online game (MMOG). A positive edge directed from node $i$ to node $j$ indicates that user $i$ perceives user $j$ positively; conversely, a negative edge directed from node $i$ to node $j$ indicates that user $i$ perceives user $j$ negatively~\cite{szell2010multirelational}.\\

\paragraph{Honduras villages dataset.} This dataset collects information about the relationships among the citizens of eleven, rural villages in the Copan Province (Western Honduras). The sample population included people older than twelve years to whom questions like `Who do you spend your free time with?', `Who is your closest friend?', `Who do you discuss personal matters with?' and `Who are the people with whom you do not get along well?' were asked (either via the Trellis software platform, available at \texttt{trellis.yale.edu}, or via tablet-based surveys, administered in a face-to-face fashion). While the answers to the first, three questions induce a positive edge, the answer to the fourth question induces a negative edge~\cite{isakov2019structure}.\\

\paragraph{Spanish high schools dataset.} This dataset collects information about the relationships among the students of thirteen high schools, three of which located in Madrid and ten of which located in Andalucia. Students were asked to name the other students they had a connection with and rate each relationship with a value ranging from $-2$ to $+2$. Since the data describe a weighted configuration, we binarised it by considering each, positive weight as a $+1$ and each, negative weight as a $-1$. As three schools are characterised by two, disconnected components and one school is characterised by four, disconnected components, we end up with nineteen, different snapshots.

\section{VI. Patterns of reciprocity in binary, directed, signed networks}

\paragraph{MMOG dataset.} Let us start commenting the empirical patterns characterising the MMOG dataset (see also Table~\ref{tab:1} for a summary of descriptive statistics). Overall, these networks are very sparse but very reciprocated: more specifically, $80\%-90\%$ of the edges come in positive, reciprocal dyads, the vast majority of the remaining ones coming in positive, single dyads: as $s^\leftrightarrow_+$ confirms, $99\%$ of the reciprocal edges is organised in dyads with a $+/+$ signature. Interestingly, there are more negative, single dyads than negative, reciprocal dyads: more quantitatively, $L^\leftrightarrow_-/L^-\simeq 0.25$, i.e. only a quarter of the negative edges appear in negative, reciprocal dyads. Lastly, it is very unlikely to observe reciprocal dyads with discordant signs.

The observations above point out the existence of a strong asymmetry between positive and negative relationships, only partly confirmed when our dataset is observed through the lenses of statistics. Figure~\ref{fig:4} shows the evolution of the aforementioned $z$-scores, i.e. $z_{L^\leftrightarrow_+}$, $z_{L^\leftrightarrow_-}$, $z_{L^\leftrightarrow_\pm}$, $z_{L^\rightarrow_+}$ and $z_{L^\rightarrow_-}$. The SDRGM and the SDCM return the same picture, i.e. $z_{L^\leftrightarrow_+}>z_{L^\leftrightarrow_-}>z_{L^\leftrightarrow_\pm}\simeq z_{L^\rightarrow_-}\simeq0>z_{L^\rightarrow_+}$: in words, the reciprocal dyads with concordant signs are over-represented in the data while the single, positive dyads are under-represented. Overall, these results point out the tendency of pairs of nodes to establish reciprocal relationships having the same sign - irrespectively from the nature of the latter one - under free-topology benchmarks.

The SDRGM-FT and the SDCM-FT, instead, return a different picture: the quantities $z_{L^\leftrightarrow_+}$, $z_{L^\leftrightarrow_-}$ and $z_{L^\rightarrow_-}$ are, in fact, positive and of comparable magnitude while $0>z_{L^\leftrightarrow_\pm}>z_{L^\rightarrow_+}$: in words, the reciprocal dyads with concordant signs and the single, negative dyads are over-represented in the data while the reciprocal dyads with discordant signs and the single, positive dyads are under-represented. Overall, these results point out the tendency of pairs of nodes to establish reciprocal relationships having the same sign - irrespectively from the nature of the latter one - as well as single, negative dyads under fixed-topology benchmarks.\\

\begin{table*}[t!]
\centering
\begin{tabular}{c||c|c|c|c||c|c||c|c|c|c|c||c|c|c}
\hline
\textit{Honduras villages} & $N$ & $L$ & $L^+$ & $L^-$ & $c$ & $r$ & $r^\leftrightarrow_+$ & $r^\leftrightarrow_-$ & $r^\leftrightarrow_\pm$ & $r^\rightarrow_+$ & $r^\rightarrow_-$ & $s^\leftrightarrow_+$ & $s^\leftrightarrow_-$ & $s^\leftrightarrow_\pm$ \\
\hline\hline
\textit{A} & 148 & 1423 & 1242 & 181 & 0.07 & 0.36 & 0.33 & 0.013 & 0.02 & 0.53 & 0.10 & 0.91 & 0.035 & 0.054 \\
\hline 
\textit{B} & 101 & 473 & 444 & 29 & 0.05 & 0.35 & 0.34 & 0.004 & 0 & 0.60 & 0.06 & 0.99 & 0.012 & 0 \\
\hline 
\textit{C} & 59 & 388 & 348 & 40 & 0.01 & 0.36 & 0.35 & 0.015 & 0 & 0.55 & 0.09 & 0.96 & 0.042 & 0 \\
\hline 
\textit{D} & 113 & 363 & 306 & 57 & 0.03 & 0.31 & 0.29 & 0.011 & 0.011 & 0.55 & 0.14 & 0.93 & 0.036 & 0.036 \\
\hline 
\textit{E} & 199 & 652 & 602 & 50 & 0.02 & 0.34 & 0.33 & 0 & 0.003 & 0.59 & 0.08 & 0.99 & 0 & 0.009 \\
\hline 
\textit{F} & 101 & 417 & 353 & 64 & 0.04 &  0.32 & 0.31 & 0.005 & 0.005 & 0.54 & 0.15 & 0.97 & 0.015 & 0.015 \\
\hline 
\textit{G} & 233 & 1070 & 924 & 146 & 0.02 & 0.34 & 0.33 & 0.008 & 0.004 & 0.53 & 0.13 & 0.97 & 0.022 & 0.011 \\
\hline 
\textit{H} & 99 & 582 & 507 & 75 & 0.06 & 0.38 & 0.37 & 0.003 & 0.007 & 0.50 & 0.12 & 0.97 & 0.009 & 0.018 \\
\hline 
\textit{I} & 98 & 321 & 285 & 36 & 0.03 & 0.39 & 0.38 & 0 & 0.006 & 0.50 & 0.11 & 0.98 & 0 & 0.016 \\
\hline 
\textit{J} & 118 & 715 & 535 & 180 & 0.05 & 0.27  & 0.24 & 0.017 & 0.014 & 0.50 & 0.23 & 0.89 & 0.062 & 0.052 \\
\hline 
\textit{K} & 206 & 890 & 708 & 182 & 0.02 & 0.32 & 0.29 & 0.016 & 0.018 & 0.50 & 0.18 & 0.89 & 0.049 & 0.056 \\
\hline
\end{tabular}
\caption{Descriptive statistics of the largest connected component of each of the eleven, rural villages composing our dataset~\cite{isakov2019structure}. The table shows the total number of nodes, $N$, the total number of edges, $L$, the total number of positive and negative edges, $L^+$ and $L^-$, the edge density, $c=L/N(N-1)$, the percentage of reciprocated edges, $r=L^{\leftrightarrow}/L$, and the empirical value of the reciprocity measures that we have introduced. Overall, these networks are very sparse and only moderately reciprocated: still, the vast majority of the existing dyads has a $+/+$ signature.}
\label{tab:2}
\end{table*}

\begin{figure*}[t!]
\centering
\includegraphics[width=\textwidth]{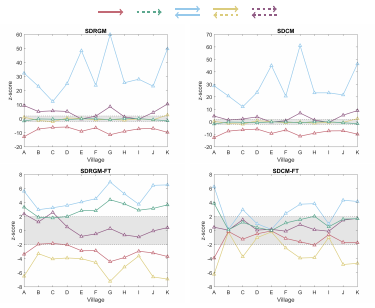}
\caption{Top panels: evolution of the $z$-scores of our dyadic motifs, under free-topology benchmarks (the SDRGM, on the left, and the SDCM, on the right), for the Honduras villages dataset. The reciprocal dyads with concordant signs (in light-blue and purple) are over-represented in the data while the single, positive dyads (in red) are under-represented in the data. Bottom panels: evolution of the $z$-scores of our dyadic motifs, under fixed-topology benchmarks (i.e. the SDRGM-FT and the SDCM-FT), for the Honduras villages dataset. While the explanatory power of the fixed-topology benchmarks is larger than that of their free-topology counterparts, the reciprocal dyads with a $+/+$ signature (in light-blue) and the single, negative dyads (in green) are over-represented in the data while the reciprocal dyads with discordant signs (in yellow) and the single, positive dyads (in red) are under-represented in the data. While the free-topology benchmarks favour the directed version of the balance theory, the fixed-topology benchmarks lead to contradictory results. The gray area corresponds to $z\in[-2,+2]$.}
\label{fig:5}
\end{figure*}

\paragraph{Honduras villages dataset.} Let us, now, comment the empirical patterns characterising the Honduras villages dataset (see also Table~\ref{tab:2} for a summary of descriptive statistics). Overall, these networks are very sparse and only moderately reciprocated: in fact, (only) one third of the edges has a companion pointing in the opposite direction. More specifically, $25\%-35\%$ of the edges come in positive, reciprocal dyads while more than half of the edges comes in positive, single dyads and more than $10\%$ of the edges comes in negative, single dyads: as $s^\leftrightarrow_+$ reveals, more than $95\%$ of the reciprocal edges is organised in dyads with a $+/+$ signature. Interestingly, there are many more negative, single dyads than negative, reciprocal dyads: more quantitatively, $L^\leftrightarrow_-/L^-\simeq 0.05$, i.e. only $5\%$ of the negative edges appear in negative, reciprocal dyads. Lastly, it is very unlikely to observe reciprocal dyads with discordant signs.

As for the MMOG dataset, a strong asymmetry between positive and negative relationships is present. Figure~\ref{fig:5} shows the evolution of the aforementioned $z$-scores: again, the SDRGM and the SDCM return the same picture, pointing out the tendency of pairs of nodes to establish reciprocal relationships having the same sign - irrespectively from the nature of the latter one. More quantitatively, $z_{L^\leftrightarrow_+}>z_{L^\leftrightarrow_-}>z_{L^\leftrightarrow_\pm}\simeq z_{L^\rightarrow_-}\simeq0>z_{L^\rightarrow_+}$, a result confirming that the reciprocal dyads with concordant signs are over-represented in the data while the single, positive dyads are under-represented.

The SDRGM-FT and the SDCM-FT, instead, return a different picture, pointing out the tendency of pairs of nodes to establish reciprocal relationships with a $+/+$ signature, as well as single, negative dyads. More quantitatively, $z_{L^\leftrightarrow_+}$ and $z_{L^\leftrightarrow_-}$ are, in fact, positive and of comparable magnitude while $z_{L^\rightarrow_-}\simeq0>z_{L^\leftrightarrow_\pm}>z_{L^\rightarrow_+}$, a result confirming that the reciprocal dyads with a $+/+$ signature and the single, negative dyads are over-represented in the data while the reciprocal dyads with discordant signs and the single, positive dyads are under-represented.\\

\paragraph{Spanish schools dataset.} Lastly, let us comment on the empirical patterns characterising the Spanish schools dataset (see also Table~\ref{tab:3} for a summary of descriptive statistics). Overall, the density of connections of these networks varies over one order of magnitude, ranging from $0.02$ to $0.2$; the reciprocity, instead, varies from $0.3$ to $0.5$. More specifically, $25\%-50\%$ of the edges come in positive, reciprocal dyads while the same percentage of edges comes in positive, single dyads and $10\%-30\%$ of the edges comes in negative, single dyads: as $s^\leftrightarrow_+$ reveals, $80\%-90\%$ of the reciprocal edges is organised in dyads with a $+/+$ signature. Interestingly, there are many more negative, single dyads than negative, reciprocal dyads: more quantitatively, $L^\leftrightarrow_-/L^-$ ranges from $0.1$ to $0.2$, i.e. only $10\%-20\%$ of the negative edges appear in negative, reciprocal dyads. Lastly, it is quite unlikely to observe reciprocal dyads with discordant signs - although more likely than observing negative, reciprocal dyads, for some of the snapshots.

\begin{figure*}[t!]
\centering
\includegraphics[width=\textwidth]{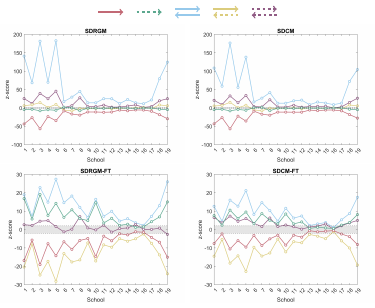}
\caption{Top panels: evolution of the $z$-scores of our dyadic motifs, under free-topology benchmarks (the SDRGM, on the left, and the SDCM, on the right), for the Spanish schools dataset. The reciprocal dyads with concordant signs (in light-blue and purple) are over-represented in the data while the single, positive dyads (in red) are under-represented in the data. Bottom panels: evolution of the $z$-scores of our dyadic motifs, under fixed-topology benchmarks (i.e. the SDRGM-FT and the SDCM-FT), for the Spanish schools dataset. The reciprocal dyads with a $+/+$ signature (in light-blue) and the single, negative dyads (in green) are over-represented in the data while the reciprocal dyads with discordant signs (in yellow) and the single, positive dyads (in red) are under-represented in the data. As for the MMOG and the Honduras villages datasets, the free-topology benchmarks favour the directed version of the balance theory while the fixed-topology benchmarks lead to contradictory results. The gray area corresponds to $z\in[-2,+2]$.}
\label{fig:6}
\end{figure*}

Figure~\ref{fig:6} shows the evolution of the aforementioned $z$-scores: according to both the SDRGM and the SDCM, the reciprocal dyads with concordant signs are over-represented in the data while the single, positive dyads are under-represented. More quantitatively, $z_{L^\leftrightarrow_+}>z_{L^\leftrightarrow_-}>z_{L^\leftrightarrow_\pm}\simeq z_{L^\rightarrow_-}\simeq0>z_{L^\rightarrow_+}$.

According to both the SDRGM-FT and the SDCM-FT, instead, the reciprocal dyads with a $+/+$ signature and the single, negative dyads are over-represented in the data while the reciprocal dyads with discordant signs and the single, positive dyads are under-represented. More quantitatively, $z_{L^\leftrightarrow_+}$ and $z_{L^\leftrightarrow_-}$ are, in fact, positive and of comparable magnitude while $z_{L^\rightarrow_-}\simeq0>z_{L^\leftrightarrow_\pm}>z_{L^\rightarrow_+}$.

\section{VII. Patterns of frustration in binary, directed, signed networks}

The structural patterns characterising our datasets highlight the tendency of nodes to establish reciprocal connections - preferentially, with concordant signs. Do these findings have a meaning from the point of view of balance? According to Cartwright and Harary, this question should be answered by disregarding the directionality of the connections constituting the cycles. In our opinion, such an approach is not fully convincing and some index accounting for directionality should, instead, be considered. 

Luckily, the three datasets considered in the present work have been collected in a similar way, i.e. letting users `rate' each other: as a consequence, the concept of balance better framing our results seems to be closer to the one informing the (directed extension of the) traditional balance theory than to the one informing the status theory. In other words, a larger number of positive, reciprocal dyads seems to characterise more balanced configurations.

\subsection{A. Strong or weak balance?}

The negative, reciprocal dyads, instead, play a role that is analogous to the role played by undirected triangles with all, negative edges: in other words, they may be considered either as a signature of balance\footnote{Such an interpretation is supported by references like~\cite{leskovec2010signed,evmenova2020analysis}, the rationale being that of identifying the notion of `balance' with the notion of `agreement': any two agreeing agents constitute a balanced dyad.} or not. In the first case, one speaks of \emph{weak balance}; in the second case, one speaks of \emph{strong balance}. Interestingly, the interpretation sustaining the weak balance obeys the (directed extension of the) definition according to which \emph{cycles with an even number of negative edges are balanced while cycles with an odd number of negative edges are frustrated} - although the intuition would suggest to not consider as balanced a couple of agents `hating' each other. More quantitatively, the number of directed cycles with zero and two, negative edges is quantified by $L^\leftrightarrow_+$ and $L^\leftrightarrow_-$, respectively, while the number of directed cycles with one, negative edge is quantified by $L^\leftrightarrow_\pm$ - and they are regarded as a signature of frustration irrespectively from the variant of the (directed extension of the) balance theory that is adopted.

\begin{table*}[t!]
\begin{tabular}{c||c|c|c|c||c|c||c|c|c|c|c||c|c|c}
\hline
\textit{Spanish schools}  & $N$ & $L$ & $L^+$ & $L^-$ & $c$ & $r$ & $r^\leftrightarrow_+$ & $r^\leftrightarrow_-$ & $r^\leftrightarrow_\pm$ & $r^\rightarrow_+$ & $r^\rightarrow_-$ & $s^\leftrightarrow_+$ & $s^\leftrightarrow_-$ & $s^\leftrightarrow_\pm$\\
\hline\hline 
1 & 409 & 8557 & 7302 & 1255 & 0.05 & 0.48 & 0.44 & 0.014 & 0.023 & 0.40 & 0.12 & 0.92 & 0.030 & 0.047 \\
\hline 
2 & 238 & 3755 & 3397 & 358 & 0.07 & 0.46 & 0.42 & 0.008 & 0.033 & 0.47 & 0.07 & 0.91 & 0.016 & 0.072 \\
\hline
3 & 534 & 12812 & 11024 & 1788 & 0.05 & 0.51 & 0.47 & 0.016 & 0.029 & 0.38 & 0.11 & 0.91 & 0.030 & 0.057 \\
\hline 
4 & 291 & 5152 & 3959 & 1193 & 0.06 & 0.35 & 0.29 & 0.031 & 0.026 & 0.46 & 0.19 & 0.84 & 0.089 & 0.074 \\
\hline 
5 & 702 & 11190 & 8242 & 2948 & 0.02 & 0.33 & 0.28 & 0.026 & 0.021 & 0.45 & 0.23 & 0.86 & 0.080 & 0.064 \\
\hline 
6 & 97 & 1805 & 1055 & 750 & 0.19 & 0.29 & 0.22 & 0.041 & 0.029 & 0.35 & 0.36 & 0.76 & 0.140 & 0.100 \\
\hline 
7 & 98 & 2080 & 1243 & 837 & 0.22 & 0.44 & 0.34 & 0.072 & 0.029 & 0.25 & 0.32 & 0.77 & 0.160 & 0.066 \\
\hline 
8 & 121 & 1622 & 1081 & 541 & 0.11 & 0.55 & 0.40 & 0.12 & 0.036 & 0.25 & 0.20 & 0.72 & 0.210 & 0.065 \\
\hline 
9 & 60 & 920 & 687 & 233 & 0.26 & 0.50 & 0.38 & 0.039 & 0.074 & 0.33 & 0.18 & 0.77 & 0.079 & 0.150 \\
\hline 
10 & 186 & 3197 & 2470 & 727 & 0.09 & 0.51 & 0.45 & 0.026 & 0.032 & 0.31 & 0.18 & 0.88 & 0.050 & 0.064 \\
\hline 
11 & 145 & 2229 & 1844 & 385 & 0.11 & 0.31 & 0.27 & 0.016 & 0.027 & 0.54 & 0.14 & 0.86 & 0.052 & 0.086 \\
\hline 
12 & 105 & 1311 & 975 & 336 & 0.12 & 0.35 & 0.31 & 0.015 & 0.027 & 0.42 & 0.23 & 0.88 & 0.043 & 0.078 \\
\hline 
13 & 91 & 935 & 791 & 144 & 0.11 & 0.27 & 0.24 & 0.009 & 0.019 & 0.60 & 0.14 &  0.90 & 0.032 & 0.072 \\
\hline 
14 & 132 & 1035 & 855 & 180 & 0.06 & 0.27 & 0.25 & 0.012 & 0.013 & 0.57 & 0.16 & 0.91 & 0.042 & 0.049 \\
\hline 
15 & 89 & 593 & 467 & 126 & 0.08 & 0.28 & 0.08 & 0.230 & 0.030 & 0.02 & 0.55 & 0.17 & 0.810 & 0.110 \\
\hline
16 & 66 & 448 & 401 & 47 & 0.10 & 0.34 & 0.31 & 0.005 & 0.270 & 0.57 & 0.09 & 0.91 & 0.013 & 0.079 \\
\hline 
17 & 124 & 1365 & 1097 & 268 & 0.09 & 0.29 & 0.26 & 0.012 & 0.019 & 0.53 & 0.18 & 0.89 & 0.040 & 0.065 \\
\hline 
18 & 391 & 4511 & 3472 & 1039 & 0.03 & 0.29 & 0.24 & 0.017 & 0.028 & 0.51 & 0.20 & 0.84 & 0.060 & 0.098 \\
\hline
19 & 458 & 7028 & 4701 & 2327 & 0.03 & 0.33 & 0.27 & 0.028 & 0.028 & 0.38 & 0.29 & 0.83 & 0.087 & 0.084 \\
\hline
\end{tabular}
\caption{Descriptive statistics of the nineteen snapshots of the Spanish schools dataset~\cite{ruiz2023triadic}. The table shows the total number of nodes, $N$, the total number of edges, $L$, the total number of positive and negative edges, $L^+$ and $L^-$, the edge density, $c=L/N(N-1)$, the percentage of reciprocated edges, $r=L^{\leftrightarrow}/L$, and the empirical value of the reciprocity measures that we have introduced. Overall, the density of connections of these networks varies over one order of magnitude while the reciprocity ranges from $0.3$ to $0.5$. As for the MMOG and the Honduras villages datasets, the vast majority of the existing dyads has a $+/+$ signature.}
\label{tab:3}
\end{table*}

\subsection{B. Free-topology or fixed-topology benchmarks?}

Analogously to what has been noticed in~\cite{gallo2024testing}, different benchmarks provide different answers to the question about the balancedness of a given directed, signed network. More specifically, free-topology benchmarks such as the SDRGM and the SDCM point out that the reciprocal dyads with concordant signs are over-represented in the data: as a consequence, \emph{our, three datasets are weakly balanced, if free-topology benchmarks are adopted.}\\

On the contrary, fixed-topology benchmarks such as the SDRGM-FT and the SDCM-FT do not lead to the same conclusions, as they point out that the negative, single dyads are over-represented as well. Should these dyads be considered? If so, how? The answer to the first question should be, in our opinion, positive, for (at least) two reasons: \emph{i)} the percentage of reciprocal links can be small, hence ignoring single dyads would imply ignoring a substantial portion of a directed, signed configuration; \emph{ii)} the magnitude of the $z$-scores of the negative, single dyads is comparable with the magnitude of the $z$-scores of the reciprocal dyads with concordant signs. What about the answer to the second question? The way our data have been collected suggests that the absence of an edge should be regarded as a lack of interest in reciprocating the interaction, whether positive or negative: as a consequence, one is led to consider the single dyads as a signature of frustration. If this is the case, however, the (directed extension of the) balance theory leads to contradictory results, as \emph{both} balanced \emph{and} frustrated patterns are over-represented in the data: as a consequence, \emph{no univocal conclusions can be drawn if fixed-topology benchmarks are adopted.}

\subsection{C. A coarser theory of balance?}

\begin{figure*}[t!]
\centering
\includegraphics[width=\textwidth]{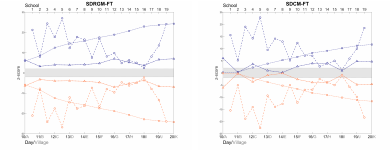}
\caption{Evolution of the $z$-scores of the coarser dyadic motifs, under fixed-topology benchmarks (i.e. the SDRGM-FT and the SDCM-FT), for the MMOG, Honduras villages and Spanish schools datasets. The reciprocal dyads with concordant signs (i.e. $N_\text{B}=L^\leftrightarrow_++L^\leftrightarrow_-$, in blue) are over-represented in the data while the reciprocal dyads with discordant signs and the single dyads (i.e. $N_\text{F}=L^\leftrightarrow_\pm+L^\rightarrow_++L^\rightarrow_-$, in orange) are under-represented in the data. In words, balance seems to be recovered at a `coarser' level. The gray area corresponds to $z\in[-2,+2]$.}
\label{fig:7}
\end{figure*}

As previously highlighted, while classifying reciprocal patterns is relatively straightforward, classifying single dyads is more problematic. If, however, we consider the two $z$-scores reading

\begin{equation}
z_\text{B}=\frac{N_\text{B}-\langle N_\text{B}\rangle}{\sigma[N_\text{B}]},
\end{equation}
where $N_\text{B}=L^\leftrightarrow_++L^\leftrightarrow_-$, and

\begin{equation}
z_\text{F}=\frac{N_\text{F}-\langle N_\text{F}\rangle}{\sigma[N_\text{F}]},
\end{equation}
where $N_\text{F}=L^\leftrightarrow_\pm+L^\rightarrow_++L^\rightarrow_-$, the contradictory patterns disappear and balance is recovered at a `coarser' level: this is evident upon looking at Figure~\ref{fig:7}, showing that balanced patterns are, now, over-represented in the data and frustrated patterns are, now, under-represented in the data - consistently across all datasets considered in the present work.

\section{VIII. Discussion}

Extending the notion of balance to directed networks is absolutely not trivial. Although such a topic has been already addressed in several papers, the latter ones have focused on triangular motifs, disregarding the information brought by the dyadic ones. Our paper represents an attempt to fill this gap, by exploring the role played by reciprocity in shaping directed, signed networks as well as understand its relationship with the notion of frustration. Our results point out the tendency of pairs of nodes to establish reciprocal relationships having the same sign and avoid reciprocal relationships having different signs. Is this a signature of balance or a signature of frustration?

A non-ambigouous answer can be provided only after having clarified the way data have been collected: in fact, as stressed in~\cite{guha2004propagation}, the presence of a signed edge, directed from a node to another, can be interpreted in several ways, depending on the `intention' pushing the source node to bind with the others. In other words, a positive edge from $i$ to $j$ can either mean `node $j$ is a friend of mine' - and the most suitable framework to interpret the results would be the one of the (directed extension of the) balance theory - or `the status of node $j$ is higher than mine' - and the most suitable framework to interpret the results would be the one of the status theory.

As the way our datasets have been constructed support the first hypothesis, we need to interpret reciprocal and single dyads in the light of the (directed extension of the) balance theory. A first route would prescribe to consider only cyclic dyads, the one with discordant signs being a signature of frustration and the ones with concordant signs being a signature of balance. At a finer level, one may even distinguish a weak, directed extension, considering both $+/+$ and $-/-$ dyads as balanced, from a strong, directed extension, solely considering $+/+$ dyads as balanced. A second route would prescribe to consider single dyads as well. From this perspective, the literature has basically ignored the problem, just focusing on the reciprocal subgraph: while this may represent an acceptable solution for the analysis of configurations with a large number of edges having a `companion' pointing in the opposite direction\footnote{Observing a large reciprocity usually leads to the conclusion that specifying the direction of the edges is, after all, unnecessary: see~\cite{garlaschelli2004patterns,Fagiolo2006DirectedOU,Fagiolo2006ClusteringIC}.}, it is no longer so when $r$ is small. A choice like this may, in fact, severely bias the analysis: as also reported in~\cite{leskovec2010signed}, the statistical evidence gained by the (directed extension of the) balance theory is larger than the one gained by the status theory \emph{on the relatively small subgraph induced by reciprocal edges}. 

Although such a framework may be adopted to interpret the results of the analysis carried out on the MMOG dataset, its use would be much less justifiable to interpret the results of the analyses carried out on the Honduras villages and Spanish schools datasets: those results call for a finer definition of balance, accounting for the meaning of single dyads. Should they be interpreted as a signature of frustration, however, the (directed extensions of the) balance theory may lead to inconsistencies preventing us from drawing univocal conclusions.

A possible way out has been proposed in~\cite{leskovec2010predicting,leskovec2010signed,evmenova2020analysis,evmenova2020structural} where it has been noticed that a substantial difference exists between single and reciprocal connections and the adoption of \emph{different} theories to explain the patterns characterising \emph{different} portions of the \emph{same} directed, signed network has been proposed - more specifically, the balance theory for the subgraph induced by reciprocal edges and the status theory for the remaining portion of the network. As such an approach would assume the \emph{simultaneous} presence of \emph{different} - if not \emph{contradictory} - processes behind the formation of a directed, signed configuration, we believe it not to represent the most suitable explanation for the emergence of certain, empirical patterns - that, instead, call for a more comprehensive analysis of directed connections in the contest of social theories.

\section{IX. Conclusions}

As our analysis highlights, directed, signed networks differ from their undirected counterparts: while the latter ones are, overall, balanced~\cite{gallo2024testing}, assessing if the former ones are balanced or not is much less straightforward. In our opinion, this is ultimately due to an incomplete formulation of the balance theory for directed networks that ignores the role played by single dyads. While such a result calls for a re-thinking of the definition of social theories for directed, signed networks, it also paves the way for a number of future research directions, such as \emph{i)} enlarging the number of datasets over which repeating the analysis proposed here; \emph{ii)} inspecting higher-order signatures of balance (e.g. triadic motifs) by employing a benchmark constraining the signed reciprocity; \emph{iii)} exploring the notion of balance at the mesoscopic level by extending the approach pursued in~\cite{gallo2024assessing}.

\section{Data availability}

The Honduras villages dataset is described in~\cite{isakov2019structure} and can be found at the address \url{https://github.com/NDS-VU/signed-network-datasets?tab=readme-ov-file#network-village}. The Spanish schools dataset is described in~\cite{ruiz2023triadic} and is publicly available at the address \url{https://zenodo.org/records/7647000#.Y-5eDtLMJH4}. The MMOG dataset, described in~\cite{szell2010multirelational}, is subject to proprietary restrictions and cannot be shared. 

\section{Acknowledgements}

DG and TS acknowledge support from SoBigData.it that receives funding from European Union – NextGenerationEU – National Recovery and Resilience Plan (Piano Nazionale di Ripresa e Resilienza, PNRR) – Project: `SoBigData.it – Strengthening the Italian RI for Social Mining and Big Data Analytics' - Prot. IR0000013 – Avviso n. 3264 del 28/12/2021. DG also acknowledges support from the Dutch Econophysics Foundation (Stichting Econophysics, Leiden, the Netherlands) and the Netherlands Organization for Scientific Research (NWO/OCW). RL acknowledges support from the EPSRC grants n. EP/V013068/1 and EP/V03474X/1.

We thank Michael Szell for sharing the Pardus dataset employed for the present analysis.

\bibliography{bibmain}

\clearpage

\onecolumngrid

\hypertarget{AppA}{}
\section{Appendix A.\\Representing binary, directed, signed networks}\label{AppA}

The three functions $a_{ij}^-=[a_{ij}=-1]$, $a_{ij}^0=[a_{ij}=0]$ and $a_{ij}^+=[a_{ij}=+1]$ have been defined via the Iverson's brackets notation. Iverson's brackets work in a way that is reminiscent of the Heaviside step function, $\Theta[x]=[x>0]$; in fact,

\begin{equation}
a_{ij}^-=[a_{ij}=-1]=\begin{dcases}
1, & \text{if}\quad a_{ij}=-1\\
0, & \text{if}\quad a_{ij}=0,+1
\end{dcases}
\end{equation}
(i.e. $a_{ij}^-=1$ if $a_{ij}<0$ and zero otherwise),

\begin{equation}
a_{ij}^0=[a_{ij}=0]=\begin{dcases}
1, & \text{if}\quad a_{ij}=0\\
0, & \text{if}\quad a_{ij}=-1,+1
\end{dcases}
\end{equation}
(i.e. $a_{ij}^0=1$ if $a_{ij}=0$ and zero otherwise),

\begin{equation}
a_{ij}^+=[a_{ij}=+1]=\begin{dcases}
1, & \text{if}\quad a_{ij}=+1\\
0, & \text{if}\quad a_{ij}=-1,0
\end{dcases}
\end{equation}
(i.e. $a_{ij}^+=1$ if $a_{ij}>0$ and zero otherwise). The matrices $\mathbf A^+\equiv\{a_{ij}^+\}_{i,j}^N$ and $\mathbf A^-\equiv\{a_{ij}^-\}_{i,j}^N$, thus, remain naturally defined and induce the relationships 

\begin{align}
\mathbf{A}&=\mathbf{A^+}-\mathbf{A^-},
\end{align}
i.e. $a_{ij}=a_{ij}^+-a_{ij}^-$, $\forall\:i\neq j$ and

\begin{align}
|\mathbf{A}|&=\mathbf{A^+}+\mathbf{A^-},
\end{align}
i.e. $|a_{ij}|=a_{ij}^++a_{ij}^-$, $\forall\:i\neq j$.

\clearpage

\hypertarget{AppB}{} 
\section{Appendix B.\\Benchmarks for binary, directed, signed networks}\label{AppB}

The generalisation of the ERG formalism to the analysis of binary, directed, signed networks can be carried out by following~\cite{park2004statistical,Squartinia}. It rests upon the constrained maximisation of Shannon entropy, i.e.

\begin{align}\label{eq:optprob}
\mathscr{L}=S[P]-\sum_{i=0}^M\theta_i[P(\mathbf{A})C_i(\mathbf{A})-\langle C_i\rangle]
\end{align}
where $S=-\sum_{\mathbf{A}\in\mathbb{A}}P(\mathbf{A})\ln P(\mathbf{A})$, $C_0\equiv\langle C_0\rangle\equiv1$ sums up the normalisation condition and the remaining $M-1$ constraints represent proper, topological properties. Such an optimisation procedure defines the expression

\begin{align}
P(\mathbf{A})=\frac{e^{-H(\mathbf{A})}}{Z}=\frac{e^{-H(\mathbf{A})}}{\sum_{\mathbf{A}\in\mathbb{A}}e^{-H(\mathbf{A})}}=\frac{e^{-\sum_{i=1}^M\theta_iC_i(\mathbf{A})}}{\sum_{\mathbf{A}\in\mathbb{A}}e^{-\sum_{i=1}^M\theta_iC_i(\mathbf{A})}}
\end{align}
that can be made explicit only after a specific set of constraints has been chosen.

\subsection{Homogeneous benchmarks for binary, directed, signed networks}

\subsubsection{Free-topology Signed Directed Random Graph Model (SDRGM)}

The Hamiltonian reading

\begin{align}
H(\mathbf{A})=\alpha L^+(\mathbf{A})+\beta L^-(\mathbf{A})
\end{align}
leads to the partition function

\begin{align}
Z&=\sum_{\mathbf{A}\in\mathbb A}e^{-H(\mathbf A)}=\sum_{\mathbf{A}\in\mathbb A}e^{-\alpha L^+(\mathbf{A})-\beta L^-(\mathbf{A})}=\sum_{\mathbf A\in\mathbb A}e^{-\sum_{i=1}^N\sum_{j(\neq i)}(\alpha a_{ij}^++\beta a_{ij}^-)}=\sum_{\mathbf A\in\mathbb A}\prod_{i=1}^N\prod_{j(\neq i)}e^{-\alpha a_{ij}^+-\beta a_{ij}^-}\nonumber\\
&=\prod_{i=1}^N\prod_{j(\neq i)}\sum_{a_{ij}=-1,0,1}e^{-\alpha a_{ij}^+-\beta a_{ij}^-}=\prod_{i=1}^N\prod_{j(\neq i)}(1+e^{-\alpha}+e^{-\beta})=(1+e^{-\alpha}+e^{-\beta})^{N(N-1)}
\end{align}
inducing the expression

\begin{align}
P_\text{SDRG}(\mathbf{A})=\frac{e^{-\alpha L^+(\mathbf{A})-\beta L^-(\mathbf{A})}}{(1+e^{-\alpha}+e^{-\beta})^{N(N-1)}}\equiv\frac{x^{L^+(\mathbf{A})}y^{L^-(\mathbf{A})}}{(1+x+y)^{N(N-1)}}\equiv(p^-)^{L^-}(p^0)^{L^0}(p^+)^{L^+}
\end{align}
having posed $p^-\equiv\frac{e^{-\beta}}{1+e^{-\alpha}+e^{-\beta}}\equiv\frac{y}{1+x+y}$, $p^0\equiv\frac{1}{1+e^{-\alpha}+e^{-\beta}}\equiv\frac{1}{1+x+y}$ and $p^+\equiv\frac{e^{-\alpha}}{1+e^{-\alpha}+e^{-\beta}}\equiv\frac{x}{1+x+y}$, where $p^+$ is the probability that any two nodes are connected by a directed, positive edge, $p^-$ is the probability that any two nodes are connected by a directed, negative edge and $p^0$ is the probability that any two nodes are no linked at all. Hence, according to the SDRGM, each entry of a directed, signed graph is a random variable following a generalised Bernoulli distribution, i.e. obeying the finite scheme

\begin{align}
a_{ij}\sim
\begin{pmatrix}
-1 & 0 & +1\\
p^- & p^0 & p^+
\end{pmatrix},\quad\forall\:i\neq j;
\end{align}
as a consequence, any graph belonging to $\mathbb{A}$ is a collection of i.i.d. random variables that obey the finite scheme

\begin{align}
\mathbf{A}\sim\bigotimes
\begin{pmatrix}
-1 & 0 & +1\\
p^- & p^0 & p^+
\end{pmatrix}
\end{align}
i.e. the directed product of the $N(N-1)$ finite schemes above.\\

The probability that, under the SDRGM, a graph has exactly $L^+$ positive edges and $L^-$ negative edges obeys the multinomial distribution

\begin{align}\label{multinomialedges}
P(L^-,L^+)=\binom{N(N-1)}{L^-,L^0,L^+}(p^-)^{L^-}(p^0)^{L^0}(p^+)^{L^+},
\end{align}
the combinatorial factor

\begin{align}\label{multinomialcoeffbis}
\binom{N(N-1)}{L^-,L^0,L^+}=\frac{(N(N-1))!}{L^-!L^0!L^+!},
\end{align}
with $L^0=N(N-1)-L=N(N-1)-(L^-+L^+)$ being the (multinomial) coefficient counting the total number of ways $L$ directed edges - $L^+$ of which are positive and $L^-$ of which are negative - can be placed among $N(N-1)$ node pairs. As a consequence, eq.~\eqref{multinomialcoeffbis} also represents the total number of directed graphs with a given number of signed edges, under the SDRGM. Naturally, $P(L^-)=\text{Bin}\left(\binom{N}{2},p^-\right)$, $P(L^0)=\text{Bin}\left(\binom{N}{2},p^0\right)$ and $P(L^+)=\text{Bin}\left(\binom{N}{2},p^+\right)$, from which it follows that the total number of expected, positive edges reads $\langle L^+\rangle=\binom{N}{2}p^+$ while the total number of expected, negative edges reads $\langle L^-\rangle=\binom{N}{2}p^-$. Obviously, $\langle L\rangle=\langle L^-\rangle+\langle L^+\rangle=\binom{N}{2}(p^-+p^+)\equiv\binom{N}{2}p$.\\

The probability that, under the SDRGM, a node has a positive in-degree exactly amounting at $h^+$ and a negative in-degree exactly amounting at $h^-$ obeys the multinomial distribution

\begin{align}\label{multinomialindeg}
P(h^-,h^+)=\binom{N-1}{h^-,h^0,h^+}(p^-)^{h^-}(p^0)^{h^0}(p^+)^{h^+};
\end{align}
analogously, the probability that a node has a positive out-degree exactly amounting at $k^+$ and a negative out-degree exactly amounting at $k^-$ obeys the multinomial distribution

\begin{align}\label{multinomialoutdeg}
P(k^-,k^+)=\binom{N-1}{k^-,k^0,k^+}(p^-)^{k^-}(p^0)^{k^0}(p^+)^{k^+},
\end{align}
with $k^0=(N-1)-k=(N-1)-(k^++k^-)$ being the (multinomial) coefficient counting the total number of ways $k$ nodes pointed by a node - $k^+$ of which via a positive edge and $k^-$ of which via a negative edge - can be selected among $N-1$ node pairs. Naturally, $P(k^-)=\text{Bin}\left(N-1,p^-\right)$, $P(k^+)=\text{Bin}\left(N-1,p^+\right)$, $P(h^-)=\text{Bin}\left(N-1,p^-\right)$ and $P(h^+)=\text{Bin}\left(N-1,p^+\right)$ from which it follows that the expected, negative out-degree reads $\langle k^-\rangle=(N-1)p^-$, the expected, positive out-degree reads $\langle k^+\rangle=(N-1)p^+$, the expected, negative in-degree reads $\langle  h^-\rangle=(N-1)p^-$ and the expected, positive in-degree reads $\langle h^+\rangle=(N-1)p^+$. Obviously, $\langle k\rangle=\langle k^-\rangle+\langle k^+\rangle=(N-1)(p^-+p^+)\equiv(N-1)p$ and $\langle h\rangle=\langle h^-\rangle+\langle h^+\rangle=(N-1)(p^-+p^+)\equiv(N-1)p$.\\

In order to determine the parameters that define the SDRGM, let us invoke the likelihood-maximisation principle. According to it, the function

\begin{align}
\mathcal{L}_\text{SDRGM}(x,y)\equiv\ln P_\text{SDRGM}(\mathbf{A}^*|x,y)=L^+(\mathbf{A}^*)\ln(x)+L^-(\mathbf{A}^*)\ln(y)-N(N-1)\ln(1+x+y)
\end{align}
has to be maximised with respect to $x$ and $y$. Upon doing so, we obtain the pair of equations

\begin{align}
\frac{\partial\mathcal{L}_\text{SDRGM}(x,y)}{\partial x}&=\frac{L^+(\mathbf{A}^*)}{x}-N(N-1)\frac{1}{1+x+y},\\
\frac{\partial\mathcal{L}_\text{SDRGM}(x,y)}{\partial y}&=\frac{L^-(\mathbf{A}^*)}{y}-N(N-1)\frac{1}{1+x+y}
\end{align}
and equating them to zero leads us to find $L^+(\mathbf{A}^*)=N(N-1)\frac{x}{1+x+y}=N(N-1)p^+=\langle L^+\rangle$ and $L^-(\mathbf{A}^*)=N(N-1)\frac{y}{1+x+y}=N(N-1)p^-=\langle L^-\rangle$, i.e. 

\begin{align}
p^+=\frac{L^+(\mathbf{A}^*)}{N(N-1)},\quad p^-=\frac{L^-(\mathbf{A}^*)}{N(N-1)}.
\end{align}

\subsubsection{Fixed-topology Signed Directed Random Graph Model (SDRGM-FT)}

Let us, now, maximise Shannon entropy, constrained to satisfy the properties $L^+(\mathbf{A})$ and $L^-(\mathbf{A})$, on a network whose topology is fixed. The Hamiltonian describing such a problem still reads

\begin{equation}
H(\mathbf{A})=\alpha L^+(\mathbf{A})+\beta L^-(\mathbf{A})
\end{equation}
but induces a partition function reading

\begin{align}
Z=&\sum_{\substack{\mathbf{A}\in\mathbb{A}\\(|\mathbf{A}|=|\mathbf{A}^*|)}}e^{-H(\mathbf{A})}
=\sum_{\substack{\mathbf A\in\mathbb A\\(|\mathbf{A}|=|\mathbf{A}^*|)}}e^{-\alpha L^+(\mathbf{A})-\beta L^-(\mathbf{A})}=\sum_{\substack{\mathbf{A}\in\mathbb{A}\\(|\mathbf{A}|=|\mathbf{A}^*|)}}e^{-\sum_{i=1}^N\sum_{j({\neq i})}(\alpha a_{ij}^++\beta a_{ij}^-)}=\sum_{\substack{\mathbf{A}\in\mathbb{A}\\(|\mathbf{A}|=|\mathbf{A}^*|)}}\prod_{i=1}^N\prod_{j(\neq i)}e^{-\alpha a_{ij}^+-\beta a_{ij}^-}\nonumber\\
=&\prod_{i=1}^N\prod_{j(\neq i)}\left(\sum_{a_{ij}=-1,1}e^{-\alpha a_{ij}^+-\beta a_{ij}^-}\right)^{|a_{ij}^*|}
=\prod_{i=1}^N\prod_{j(\neq i)}(e^{-\alpha}+e^{-\beta})^{|a_{ij}^*|}=(e^{-\alpha}+e^{-\beta})^L
\end{align}
where the support of the distribution becomes the set of node pairs $i,j$, with $i\neq j$, such that $|a_{ij}^*|=1$. The expression above leads to

\begin{align}\label{probgraph}
P_\text{SDRGM-FT}(\mathbf{A})=\frac{e^{-\alpha L^+(\mathbf{A})-\beta L^-(\mathbf{A})}}{(e^{-\alpha}+e^{-\beta})^L}\equiv\frac{x^{L^+(\mathbf{A})}y^{L^-(\mathbf{A})}}{(x+y)^L}\equiv(p^-)^{L^-}(p^+)^{L^+}
\end{align}
having posed $p^-\equiv\frac{e^{-\beta}}{e^{-\alpha}+e^{-\beta}}\equiv\frac{y}{x+y}$ and $p^+\equiv\frac{e^{-\alpha}}{e^{-\alpha}+e^{-\beta}}\equiv\frac{x}{x+y}$, where $p^+$ is the probability that any two nodes are connected by a directed, positive edge and $p^-$ is the probability that any two nodes are connected by a directed, negative edge. Hence, according to the SDRGM-FT, each entry of a directed, signed graph such that $|a_{ij}^*|=1$ is a random variable following a Bernoulli distribution, i.e. obeying the finite scheme

\begin{equation}
a_{ij}\sim
\begin{pmatrix}
-1 & +1\\
p^- & p^+
\end{pmatrix},\quad\forall\:i\neq j\:|\:|a^*_{ij}|=1.
\end{equation}

The probability that, under the SDRGM-FT, a graph has exactly $L^+$ positive edges obeys the binomial distribution

\begin{equation}
P(L^+)=\binom{L}{L^+}(p^-)^{L^-}(p^+)^{L^+}=\binom{L}{L^+}(p^+)^{L^+}(1-p^+)^{L-L^+};
\end{equation}
as a consequence, the total number of expected, positive edges reads $\langle L^+\rangle=Lp^+$; analogously, for $L^-\sim\text{Bin}(L,p^-)$. Similarly, the probability that a node has a positive in-degree exactly amounting at $h^+$ obeys the binomial distribution

\begin{align}
P(h^+)=\binom{h}{h^+}(p^-)^{h^-}(p^+)^{h^+}=\binom{h}{h^+}(p^+)^{h^+}(1-p^-)^{h-h^+}
\end{align}
and the probability that a node has a positive out-degree exactly amounting at $k^+$ obeys the binomial distribution

\begin{align}
P(k^+)=\binom{k}{ k^+}(p^-)^{k^-}(p^+)^{k^+}=\binom{k}{k^+}(p^+)^{k^+}(1-p^-)^{k-k^+};
\end{align}
analogously, $h^-\sim\text{Bin}(h,p^-)$ and $k^-\sim\text{Bin}(k,p^-)$. As a consequence, the expected, positive and negative in-degrees read $\langle h^+\rangle=hp^+$ and $\langle h^-\rangle=hp^-$ while the expected, positive and negative out-degrees read $\langle k^+\rangle=kp^+$ and $\langle k^-\rangle=kp^-$.\\

In order to determine the parameters that define the SDRGM-FT, let us invoke the likelihood-maximisation principle again. According to it, the function 

\begin{equation}
\mathcal{L}_\text{SDRGM-FT}(x,y)\equiv\ln P_\text{SDRGM-FT}(\mathbf{A}^*|x,y)=L^+(\mathbf{A}^*)\ln(x)+L^-(\mathbf{A}^*)\ln(y)-L(\mathbf{A}^*)\ln(x+y)
\end{equation}
has to be maximized with respect to $x$ and $y$. Upon doing so, we obtain the pair of equations

\begin{align}
\frac{\partial\mathcal{L}_\text{SDRGM-FT}(x,y)}{\partial x}&=\frac{L^+(\mathbf{A}^*)}{x}-\frac{L(\mathbf{A}^*)}{x+y},\\
\frac{\partial\mathcal{L}_\text{SDRGM-FT}(x,y)}{\partial y}&=\frac{L^-(\mathbf{A}^*)}{y}-\frac{L(\mathbf{A}^*)}{x+y}
\end{align}
and equating them to zero leads us to find $L^+(\mathbf{A}^*)=L(\mathbf{A}^*)\frac{x}{x+y}=L(\mathbf{A}^*)p^+=\langle L^+\rangle$ and $L^-(\mathbf{A}^*)=L(\mathbf{A}^*)\frac{y}{x+y}=L(\mathbf{A}^*)p^-=\langle L^-\rangle$, i.e. 

\begin{align}
p^+=\frac{L^+(\mathbf{A}^*)}{L(\mathbf{A}^*)},\quad p^-=\frac{L^-(\mathbf{A}^*)}{L(\mathbf{A}^*)}.
\end{align}

\subsubsection{An alternative derivation of the SDRGM-FT}

Let us explicitly notice that $p^+$ can be re-written as follows:

$$
p^+=\frac{x}{x+y}=\frac{x/y}{x/y+1}\equiv\frac{z}{1+z};
$$
beside reducing the number of unknowns, such a relationship sheds light on the role played by the constraints $L^+(\mathbf{A})$ and $L^-(\mathbf{A})$. This becomes evident upon considering that $z=e^{-(\alpha-\beta)}$, an expression suggesting that the only relevant multiplier is $\alpha-\beta$. This is indeed the case: upon re-writing $H(\mathbf{A})$ as

\begin{equation}
H(\mathbf{A})=\alpha L^+(\mathbf{A})+\beta L^-(\mathbf{A})=\alpha L^+(\mathbf{A})+\beta [L(\mathbf{A})-L^+(\mathbf{A})]=(\alpha-\beta)L^+(\mathbf{A})-\beta L(\mathbf{A})
\end{equation}
the calculations lead to the expression above~\cite{hao2024proper}.

\subsection{Heterogeneous benchmarks for binary, directed, signed networks}

\subsubsection{Free-topology Signed Directed Configuration Model (SDCM)}

The Hamiltonian reading 

\begin{align}
H(\mathbf{A})=\sum_{i=1}^N[\alpha_ik_i^+(\mathbf{A})+\beta_ik_i^-(\mathbf{A})+\gamma_ih_i^+(\mathbf{A})+\delta_ih_i^-(\mathbf{A})]
\end{align}
leads to the partition function

\begin{align}
Z&=\sum_{\mathbf{A}\in\mathbb A}e^{-H(\mathbf{A})}=\sum_{\mathbf{A}\in\mathbb{A}}e^{-\sum_{i=1}^N[\alpha_ik_i^+(\mathbf{A})+\beta_ik_i^-(\mathbf{A})+\gamma_ih_i^+(\mathbf{A})+\delta_ih_i^-(\mathbf{A})]}\nonumber\\
&=\sum_{\mathbf{A}\in\mathbb{A}}e^{-\sum_{i=1}^N\sum_{j(\neq i)}[(\alpha_i+\gamma_j)a_{ij}^++(\beta_i+\delta_j)a_{ij}^-]}=\sum_{\mathbf{A}\in\mathbb{A}}\prod_{i=1}^N\prod_{j(\neq i)}e^{-(\alpha_i+\gamma_j)a_{ij}^+-(\beta_i+\delta_j)a_{ij}^-}\nonumber\\
&=\prod_{i=1}^N\prod_{j(\neq i)}\sum_{a_{ij}=-1,0,1}e^{-(\alpha_i+\gamma_j)a_{ij}^+-(\beta_i+\delta_j)a_{ij}^-}=\prod_{i=1}^N\prod_{j(\neq i)}(1+e^{-(\alpha_i+\gamma_j)}+e^{-(\beta_i+\delta_j)})
\end{align}
inducing the expression

\begin{align}
P_\text{SDCM}(\mathbf{A})&=\frac{e^{-\sum_{i=1}^N[\alpha_i  k_i^+(\mathbf{A})+\beta_i k_i^-(\mathbf{A})+\gamma_i  h_i^+(\mathbf{A})+\delta_i h_i^-(\mathbf{A})]}}{\prod_{i=1}^N\prod_{j(\neq i)}(1+e^{-(\alpha_i+\gamma_j)}+e^{-(\beta_i+\delta_j)})}\nonumber\\
&\equiv\frac{\prod_{i=1}^Nx_i^{k_i^+(\mathbf{A})}y_i^{k_i^-(\mathbf{A})}z_i^{h_i^+(\mathbf{A})}w_i^{h_i^-(\mathbf{A})}}{\prod_{i=1}^N\prod_{j(\neq i)}(1+x_iz_j+y_iw_j)}\nonumber\\
&\equiv\prod_{i=1}^N\prod_{j(\neq i)}(p_{ij}^-)^{a_{ij}^-}(p_{ij}^0)^{a_{ij}^0}(p_{ij}^+)^{a_{ij}^+}
\end{align}
having posed $p_{ij}^-\equiv\frac{e^{-(\beta_i+\delta_j)}}{1+e^{-(\alpha_i+\gamma_j)}+e^{-(\beta_i+\delta_j)}}\equiv\frac{y_iw_j}{1+x_iz_j+y_iw_j}$, $p_{ij}^0\equiv\frac{1}{1+e^{-(\alpha_i+\gamma_j)}+e^{-(\beta_i+\delta_j)}}\equiv\frac{1}{1+x_iz_j+y_iw_j}$ and $p_{ij}^+\equiv\frac{e^{-(\alpha_i+\gamma_j)}}{1+e^{-(\alpha_i+\gamma_j)}+e^{-(\beta_i+\delta_j)}}\equiv\frac{x_iz_j}{1+x_iz_j+y_iw_j}$, where $p_{ij}^+$ is the probability that node $i$ points node $j$ via a positive edge, $p_{ij}^-$ is the probability that node $i$ points node $j$ via a negative edge and $p_{ij}^0$ is the probability that nodes $i$ and $j$ are no linked at all. Hence, according to the SDCM, the generic entry of a directed, signed graph is a random variable following a generalised Bernoulli distribution, i.e. obeying the finite scheme

\begin{equation}
a_{ij}\sim
\begin{pmatrix}
-1 & 0 & +1\\
p_{ij}^- & p_{ij}^0 & p_{ij}^+
\end{pmatrix},\quad\forall\:i\neq j;
\end{equation}
as a consequence, any graph belonging to $\mathbb{A}$ is a collection of i.i.d. random variables that obey the finite scheme

\begin{equation}
\mathbf{A}\sim\bigotimes
\begin{pmatrix}
-1 & 0 & +1\\
p_{ij}^- & p_{ij}^0 & p_{ij}^+
\end{pmatrix}
\end{equation}
i.e. the directed product of the $N(N-1)$ finite schemes above.\\

In the case of the SDCM, $L^+$ and $L^-$ are random variables obeying Poisson-Binomial distributions, i.e. $L^+\sim\text{PoissBin}\left(N(N-1),\{p_{ij}^+\}_{i,j=1}^N\right)$ and $L^-\sim\text{PoissBin}\left(N(N-1),\{p_{ij}^-\}_{i,j=1}^N\right)$. Similarly, $k_i^+\sim\text{PoissBin}\left(N-1,\{p_{ij}^+\}_{j=1}^N\right)$, $k_i^-\sim\text{PoissBin}\left(N-1,\{p_{ij}^-\}_{j=1}^N\right)$, $h_i^+\sim\text{PoissBin}\left(N-1,\{p_{ji}^+\}_{j=1}^N\right)$ and $h_i^-\sim\text{PoissBin}\left(N-1,\{p_{ji}^-\}_{j=1}^N\right)$. Hence, the total number of expected, positive, directed edges reads $\langle L^+\rangle=\sum_{i=1}^N\sum_{j(\neq i)=1}^Np_{ij}^+$ while the total number of expected, negative, directed edges reads $\langle L^-\rangle=\sum_{i=1}^N\sum_{j(\neq 
i)=1}^Np_{ij}^-$; analogously, $\langle k_i^+\rangle=\sum_{j(\neq i)=1}^Np_{ij}^+$, $\langle k_i^-\rangle=\sum_{j(\neq i)=1}^Np_{ij}^-$, $\langle  h_i^+\rangle=\sum_{j(\neq i)=1}^Np_{ji}^+$ and $\langle h_i^-\rangle=\sum_{j(\neq i)=1}^Np_{ji}^-$.\\

In order to determine the parameters that define the SDCM, let us invoke the likelihood-maximisation principle. According to it, the function

\begin{align}
\mathcal{L}_\text{SDCM}&(\mathbf{x},\mathbf{y},\mathbf{z},\mathbf{w})\equiv\ln P_\text{SDCM}(\mathbf{A}^*|\mathbf{x},\mathbf{y},\mathbf{z},\mathbf{w})\nonumber\\
&=\sum_{i=1}^N k_i^+(\mathbf{A}^*)\ln(x_i)+\sum_{i=1}^N k_i^-(\mathbf{A}^*)\ln(y_i)+\sum_{i=1}^N h_i^+(\mathbf{A}^*)\ln(z_i)+\sum_{i=1}^N h_i^-(\mathbf{A}^*)\ln(w_i)-\sum_{i=1}^N\sum_{j(\neq i)}\ln(1+x_iz_j+y_iw_j)
\end{align}
has to be maximized with respect to $x_i$, $y_i$, $z_i$ and $w_i$, $\forall\:i$. Upon doing so, we obtain the system of $4N$ equations

\begin{align}
\frac{\partial\mathcal{L}_\text{SDCM}(\mathbf{x},\mathbf{y},\mathbf{z},\mathbf{w})}{\partial x_i}&=\frac{ k_i^+(\mathbf{A}^*)}{x_i}-\sum_{j(\neq i)}\frac{z_j}{1+x_iz_j+y_iw_j}\quad\forall\:i,\\
\frac{\partial\mathcal{L}_\text{SDCM}(\mathbf{x},\mathbf{y},\mathbf{z},\mathbf{w})}{\partial y_i}&=\frac{ k_i^-(\mathbf{A}^*)}{y_i}-\sum_{j(\neq i)}\frac{w_j}{1+x_iz_j+y_iw_j}\quad\forall\:i,\\
\frac{\partial\mathcal{L}_\text{SDCM}(\mathbf{x},\mathbf{y},\mathbf{z},\mathbf{w})}{\partial z_i}&=\frac{ h_i^+(\mathbf{A}^*)}{z_i}-\sum_{j(\neq i)}\frac{x_j}{1+x_jz_i+y_jw_i}\quad\forall\:i,\\
\frac{\partial\mathcal{L}_\text{SDCM}(\mathbf{x},\mathbf{y},\mathbf{z},\mathbf{w})}{\partial w_i}&=\frac{ h_i^-(\mathbf{A}^*)}{w_i}-\sum_{j(\neq i)}\frac{y_j}{1+x_jz_i+y_jw_i}\quad\forall\:i;
\end{align}
equating them to zero leads us to find

\begin{align}
 k_i^+(\mathbf{A}^*)&=\sum_{j(\neq i)}\frac{x_iz_j}{1+x_iz_j+y_iw_j}=\sum_{j(\neq i)}p_{ij}^+=\langle k_i^+\rangle\quad\forall\:i,\\
 k_i^-(\mathbf{A}^*)&=\sum_{j(\neq i)}\frac{y_iw_j}{1+x_iz_j+y_iw_j}=\sum_{j(\neq i)}p_{ij}^-=\langle k_i^-\rangle\quad\forall\:i,\\
 h_i^+(\mathbf{A}^*)&=\sum_{j(\neq i)}\frac{x_jz_i}{1+x_jz_i+y_jw_i}=\sum_{j(\neq i)}p_{ji}^+=\langle h_i^+\rangle\quad\forall\:i,\\
 h_i^-(\mathbf{A}^*)&=\sum_{j(\neq i)}\frac{y_jw_i}{1+x_jz_i+y_jw_i}=\sum_{j(\neq i)}p_{ji}^-=\langle h_i^-\rangle\quad\forall\:i.
\end{align}

Although the system above can be solved only numerically, particular conditions exist under which it can be solved explicitly. They are collectively named `sparse-case approximation of the SDCM' and hold true whenever $x_i\ll1$, $y_i\ll1$, $z_i\ll1$, $w_i\ll1$, $\forall\:i$.

\subsubsection{Fixed-topology Signed Configuration Model (SDCM-FT)}

Let us, now, maximize Shannon entropy, constrained to satisfy the properties $\{k_i^+(\mathbf{A})\}_{i=1}^N$, $\{k_i^-(\mathbf{A})\}_{i=1}^N$, $\{h_i^+(\mathbf{A})\}_{i=1}^N$ and $\{h_i^-(\mathbf{A})\}_{i=1}^N$, on a network whose topology is fixed. The Hamiltonian describing such a problem still reads

\begin{align}
H(\mathbf{A})=\sum_{i=1}^N[\alpha_i k_i^+(\mathbf{A})+\beta_i k_i^-(\mathbf{A})+\gamma_i h_i^+(\mathbf{A})+\delta_i h_i^-(\mathbf{A})]
\end{align}
but induces a partition function reading 

\begin{align}
Z&=\sum_{\substack{\mathbf{A}\in\mathbb{A}\\(|\mathbf{A}|=|\mathbf{A^*}|)}}e^{-H(\mathbf{A})}=\sum_{\substack{\mathbf{A}\in\mathbb{A}\\(|\mathbf{A}|=|\mathbf{A^*}|)}}e^{-\sum_{i=1}^N[\alpha_ik_i^+(\mathbf{A})+\beta_ik_i^-(\mathbf{A})+\gamma_ih_i^+(\mathbf{A})+\delta_ih_i^-(\mathbf{A})]}\nonumber\\
&=\sum_{\substack{\mathbf{A}\in\mathbb{A}\\(|\mathbf{A}|=|\mathbf{A^*}|)}}e^{-\sum_{i=1}^N\sum_{{j(\neq i})=1}^N[(\alpha_i+\gamma_j)a_{ij}^++(\beta_i+\delta_j)a_{ij}^-]}=\sum_{\substack{\mathbf{A}\in\mathbb{A}\\(|\mathbf{A}|=|\mathbf{A^*}|)}}\prod_{i=1}^N\prod_{j(\neq i)}e^{-(\alpha_i+\gamma_j)a_{ij}^+-(\beta_i+\delta_j)a_{ij}^-}\nonumber\\
&=\prod_{i=1}^N\prod_{j(\neq i)}\left(\sum_{a_{ij}=-1,1}e^{-(\alpha_i+\gamma_j)a_{ij}^+-(\beta_i+\delta_j)a_{ij}^-}\right)^{|a^*_{ij}|}=\prod_{i=1}^N\prod_{j(\neq i)}\left(e^{-(\alpha_i+\gamma_j)}+e^{-(\beta_i+\delta_j)}\right)^{|a^*_{ij}|}.
\end{align}

The expression above induces the expression

\begin{align}
P_\text{SDCM-FT}(\mathbf{A})&=\frac{e^{-\sum_{i=1}^N[\alpha_ik_i^+(\mathbf{A})+\beta_ik_i^-(\mathbf{A})+\gamma_ih_i^+(\mathbf{A})+\delta_ih_i^-(\mathbf{A})]}}{\prod_{i=1}^N\prod_{j(\neq i)}\left(e^{-(\alpha_i+\gamma_j)}+e^{-(\beta_i+\delta_j)}\right)^{|a^*_{ij}|}}\nonumber\\
&\equiv\frac{\prod_{i=1}^Nx_i^{ k_i^+(\mathbf{A})}y_i^{k_i^-(\mathbf{A})}z_i^{h_i^+(\mathbf{A})}w_i^{h_i^-(\mathbf{A})}}{\prod_{i=1}^N\prod_{j(\neq i)}\left(x_iz_j+y_iw_j\right)^{|a^*_{ij}|}}\nonumber\\
&\equiv\prod_{i=1}^N\prod_{j(\neq i)}(p_{ij}^-)^{a_{ij}^-}(p_{ij}^+)^{a_{ij}^+}
\end{align}
having posed $p_{ij}^-\equiv\frac{e^{-(\beta_i+\delta_j)}}{e^{-(\alpha_i+\gamma_j)}+e^{-(\beta_i+\delta_j)}}\equiv\frac{y_iw_j}{x_iz_j+y_iw_j}$ and $p_{ij}^+\equiv\frac{e^{-(\alpha_i+\gamma_j)}}{e^{-(\alpha_i+\gamma_j)}+e^{-(\beta_i+\delta_j)}}\equiv\frac{x_iz_j}{x_iz_j+y_iw_j}$, where $p_{ij}^+$ is the probability that node $i$ points node $j$ via a positive edge and $p_{ij}^-$ is the probability that node $i$ points node $j$ via a negative edge. Hence, according to the SDCM-FT, the generic entry of a directed, signed graph is a random variable following a Bernoulli distribution, i.e. obeying the finite scheme

\begin{equation}
a_{ij}\sim
\begin{pmatrix}
-1 & +1\\
p_{ij}^- & p_{ij}^+
\end{pmatrix},\quad\forall\:i\neq j\:|\:|a^*_{ij}|=1.
\end{equation}

In the case of the SDCM-FT, $L^+$ and $L^-$ are random variables obeying Poisson-Binomial distributions, i.e. $L^+\sim\text{PoissBin}\left(L,\{p_{ij}^+\}_{i,j=1}^N\right)$ and $L^-\sim\text{PoissBin}\left(L,\{p_{ij}^-\}_{i,j=1}^N\right)$. Similarly, $k_i^+\sim\text{PoissBin}\left(k_i,\{p_{ij}^+\}_{j=1}^N\right)$, $k_i^-\sim\text{PoissBin}\left(k_i,\{p_{ij}^-\}_{j=1}^N\right)$, $h_i^+\sim\text{PoissBin}\left(h_i,\{p_{ij}^+\}_{j=1}^N\right)$ and $h_i^-\sim\text{PoissBin}\left( h_i,\{p_{ij}^-\}_{j=1}^N\right)$. Hence, the total number of expected, positive, directed edges reads $\langle L^+\rangle=\sum_{i=1}^N\sum_{j({\neq i})=1}^N|a_{ij}^*|p_{ij}^+$ while the total number of expected, negative, directed edges reads $\langle L^-\rangle=\sum_{i=1}^N\sum_{j({\neq i})=1}^N|a_{ji}^*|p_{ij}^-$; analogously, $\langle k_i^+\rangle=\sum_{j({\neq i})=1}^N|a_{ij}^*|p_{ij}^+$, $\langle k_i^-\rangle=\sum_{j({\neq i})=1}^N|a_{ij}^*|p_{ij}^-$, $\langle h_i^+\rangle=\sum_{j({\neq i})=1}^N|a_{ji}^*|p_{ji}^+$ and $\langle h_i^-\rangle=\sum_{j({\neq i})=1}^N|a_{ji}^*|p_{ji}^-$.\\

In order to determine the parameters that define the SDCM-FT, let us invoke the likelihood-maximisation principle again. According to it, the function

\begin{align}
&\mathcal{L}_\text{SDCM-FT}(\mathbf{x},\mathbf{y},\mathbf{z},\mathbf{w})\equiv\ln P_\text{SDCM-FT}(\mathbf{A}^*|\mathbf{x},\mathbf{y},\mathbf{z},\mathbf{w})\nonumber\\
&=\sum_{i=1}^N k_i^+(\mathbf{A}^*)\ln(x_i)+\sum_{i=1}^N k_i^-(\mathbf{A}^*)\ln(y_i)+\sum_{i=1}^N h_i^+(\mathbf{A}^*)\ln(z_i)+\sum_{i=1}^N h_i^-(\mathbf{A}^*)\ln(w_i)-\sum_{i=1}^N\sum_{j(\neq i)}|a^*_{ij}|\ln(x_iz_j+y_iw_j)
\end{align}
has to be maximized with respect to $x_i$, $y_i$, $z_i$ and $w_i$, $\forall\:i$. Upon doing so, we obtain the system of equations

\begin{align}
\frac{\partial\mathcal{L}_\text{SDCM}(\mathbf{x},\mathbf{y},\mathbf{z},\mathbf{w})}{\partial x_i}&=\frac{ k_i^+(\mathbf{A}^*)}{x_i}-\sum_{j(\neq i)}|a^*_{ij}|\frac{z_j}{x_iz_j+y_iw_j}\quad\forall\:i,\\
\frac{\partial\mathcal{L}_\text{SDCM}(\mathbf{x},\mathbf{y},\mathbf{z},\mathbf{w})}{\partial y_i}&=\frac{ k_i^-(\mathbf{A}^*)}{y_i}-\sum_{j(\neq i)}|a^*_{ij}|\frac{w_j}{x_iz_j+y_iw_j}\quad\forall\:i,\\
\frac{\partial\mathcal{L}_\text{SDCM}(\mathbf{x},\mathbf{y},\mathbf{z},\mathbf{w})}{\partial z_i}&=\frac{ h_i^+(\mathbf{A}^*)}{z_i}-\sum_{j(\neq i)}|a^*_{ji}|\frac{x_j}{x_jz_i+y_jw_i}\quad\forall\:i,\\
\frac{\partial\mathcal{L}_\text{SDCM}(\mathbf{x},\mathbf{y},\mathbf{z},\mathbf{w})}{\partial w_i}&=\frac{ h_i^-(\mathbf{A}^*)}{w_i}-\sum_{j(\neq i)}|a^*_{ji}|\frac{y_j}{x_jz_i+y_jw_i}\quad\forall\:i;
\end{align}
equating them to zero leads us to find

\begin{align}
 k_i^+(\mathbf{A}^*)&=\sum_{j(\neq i)}|a^*_{ij}|\frac{x_iz_j}{x_iz_j+y_iw_j}=\sum_{j(\neq i)}|a^*_{ij}|p_{ij}^+=\langle k_i^+\rangle\quad\forall\:i,\\
 k_i^-(\mathbf{A}^*)&=\sum_{j(\neq i)}|a^*_{ij}|\frac{y_iw_j}{x_iz_j+y_iw_j}=\sum_{j(\neq i)}|a^*_{ij}|p_{ij}^-=\langle k_i^-\rangle\quad\forall\:i,\\
 h_i^+(\mathbf{A}^*)&=\sum_{j(\neq i)}|a^*_{ji}|\frac{x_jz_i}{x_jz_i+y_jw_i}=\sum_{j(\neq i)}|a^*_{ji}|p_{ji}^+=\langle h_i^+\rangle\quad\forall\:i,\\
 h_i^-(\mathbf{A}^*)&=\sum_{j(\neq i)}|a^*_{ji}|\frac{y_jw_i}{x_jz_i+y_jw_i}=\sum_{j(\neq i)}|a^*_{ji}|p_{ji}^-=\langle h_i^-\rangle\quad\forall\:i.
\end{align}

The system above can be solved only numerically.

\subsubsection{An alternative derivation of the SDCM-FT}

Let us explicitly notice that $p_{ij}^+$ can be re-written as follows:

$$
p_{ij}^+=\frac{x_iz_j}{x_iz_j+y_iw_j}=\frac{x_iz_j/y_iw_j}{x_iz_j/y_iw_j+1}\equiv\frac{s_it_j}{1+s_it_j};
$$
beside reducing the number of unknowns, such a relationship sheds light on the role played by the constraints $k_i^+(\mathbf{A})$, $k_i^-(\mathbf{A})$, $h_i^+(\mathbf{A})$, $h_i^-(\mathbf{A})$. This becomes evident upon considering that $s_i=e^{-(\alpha_i-\gamma_j)}$ and $t_i=e^{-(\beta_i-\delta_j)}$, expressions suggesting that the only relevant multipliers are $\alpha_i-\gamma_j$ and $\beta_i-\delta_j$. This is indeed the case: upon re-writing $H(\mathbf{A})$ as

\begin{align}
H(\mathbf{A})&=\sum_{i=1}^N[\alpha_i k_i^+(\mathbf{A})+\beta_i k_i^-(\mathbf{A})+\gamma_i h_i^+(\mathbf{A})+\delta_i h_i^-(\mathbf{A})]\nonumber\\
&=\sum_{i=1}^N[\alpha_i k_i^+(\mathbf{A})+\beta_i (k_i(\mathbf{A})-k_i^+(\mathbf{A}))+\gamma_i h_i^+(\mathbf{A})+\delta_i (h_i(\mathbf{A})-h_i^+(\mathbf{A}))]\nonumber\\
&=\sum_{i=1}^N[(\alpha_i-\beta_i)k_i^+(\mathbf{A})+(\gamma_i-\delta_i)h_i^+(\mathbf{A})]+\sum_{i=1}^N[\beta_i k_i(\mathbf{A})+\delta_i h_i(\mathbf{A})]
\end{align}
the calculations lead to the expression above~\cite{hao2024proper}.

\clearpage 

\hypertarget{AppC}{}
\section{Appendix C.\\Numerical optimisation of likelihood functions}\label{AppC}

In~\cite{vallarano2021fast}, the authors solve several ERGs (defined by binary and weighted, undirected and directed constraints) by implementing and comparing three, numerical methods. In order to solve the SDCM and the SDCM-FT, here we extend the iterative recipe to our setting: more precisely, it is a fixed-point iteration aiming to estimate $\vec\theta$ such that $\vec\theta=\mathbf G(\vec\theta)$, where $\mathbf G$ is an appropriate function defined on real numbers. Such an iterative algorithm can be read as a modified version of the Karush-Kuhn-tucker (KKT) conditions and the explicit definition of each iteration can be re-written as $\vec\theta^{(n)}=\mathbf G(\vec\theta^{(n-1)})$. The freely downloadable Matlab package implementing it is named \href{https://it.mathworks.com/matlabcentral/fileexchange/167426-signed-models-for-network-analysis}{`SIMONA'}.\\

We adopt two different stopping criteria: the first one puts a condition on the Euclidean norm of the vector of differences between the values of the parameters at subsequent iterations, i.e.

\begin{align}
||\Delta\vec\theta||_2=\sqrt{\sum_{i=1}^N(\Delta\theta_i)^2}\le10^{-8};
\end{align}
the second one stops the algorithm when the maximum number of iterations (set to 1.000) is reached.\\

The accuracy of the fixed-point method in estimating the constraints has been evaluated via the \emph{maximum absolute error} (MAE) defined as $\max_{i}\{|C_i^*-\langle C_i \rangle|\}_{i=1}^N$, i.e. the infinite norm of the difference between the vector of the empirical values of the constraints, $C_i^*$, and the vector of their expected values, $\langle C_i\rangle$. In our setting, it explicitely reads

\begin{align}
\text{MAE}=\max_{i=1\dots N}\{|(k_i^+)^*-\langle k_i^+\rangle|\:,\:|(k_i^-)^*-\langle k_i^-\rangle|\:,\:|(h_i^+)^*-\langle h_i^+\rangle|\:,\:|(h_i^-)^*-\langle h_i^-\rangle|\}.
\end{align}

In order to compare the magnitude of the absolute error with the numerical value of the constraint it refers to, we also quantify the accuracy of our recipe via the \emph{maximum relative error} (MRE) defined as

\begin{align}
\text{MRE}=\max_{i=1\dots N}\left\{\frac{|(k_i^+)^*-\langle k_i^+\rangle|}{k_i^+}\:,\:\frac{|(k_i^-)^*-\langle k_i^-\rangle|}{k_i^-}\:,\:\frac{|(h_i^+)^*-\langle h_i^+\rangle|}{h_i^+}\:,\:\frac{|(h_i^-)^*-\langle h_i^-\rangle|}{h_i^-}\right\}.
\end{align}

\begin{figure}[t!]
\centering
\includegraphics[width=\textwidth]{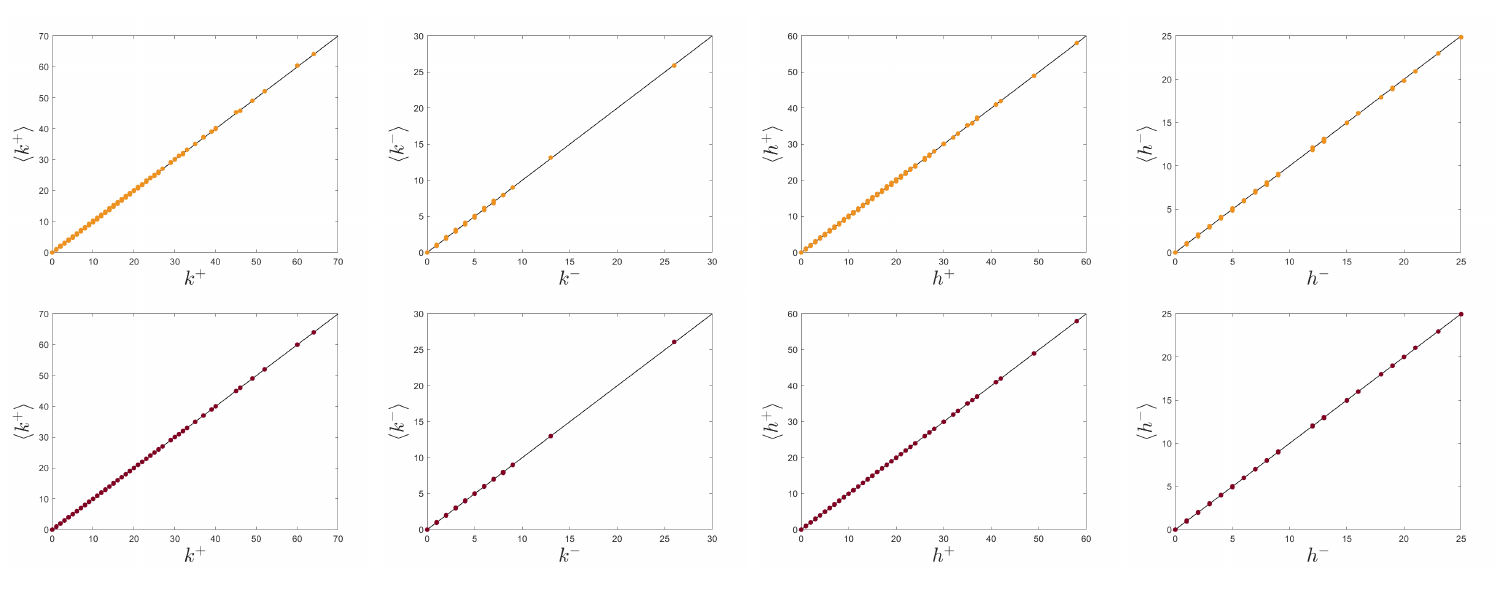}
\caption{Top panels: comparison between the empirical value and the sample average of the positive out-degree, negative out-degree, positive in-degree and negative in-degree for each node in the network, under the SDCM, on the 20th day of the MMOG dataset. Bottom panels: comparison between the empirical value and the sample average of the positive out-degree, negative out-degree, positive in-degree and negative in-degree for each node in the network, under the SDCM-FT, on the 20th day of the MMOG dataset.}
\label{fig:8}
\end{figure}

For an illustrative example of the accuracy of our method, we refer to Figure~\ref{fig:8}.

\subsection{Resolution of the SDCM}

The system of equations embodying the SDCM first-order optimality conditions can be re-written as

\begin{align}
\label{eq:iterD1}
x_i&=\frac{k_i^+(\mathbf{A^*})}{\sum_{j(\neq i)}\frac{z_j}{1+x_iz_j+y_iw_j}}\quad\forall\:i,\\
y_i&=\frac{k_i^-(\mathbf{A^*})}{\sum_{j(\neq i)}\frac{w_j}{1+x_iz_j+y_iw_j}}\quad\forall\:i,\\
z_i&=\frac{h_i^+(\mathbf{A^*})}{\sum_{j(\neq i)}\frac{x_j}{1+x_jz_i+y_jw_i}}\quad\forall\:i,\\
\label{eq:iterD4}
w_i&=\frac{h_i^-(\mathbf{A^*})}{\sum_{j(\neq i)}\frac{y_j}{1+x_jz_i+y_jw_i}}\quad\forall\:i.
\end{align}

Upon noticing that the terms on the l.h.s. of Equations \eqref{eq:iterD1}-\eqref{eq:iterD4} appear also in the corresponding r.h.s., the iterative recipe that has to be solved can be written as

\begin{align}
x_i^{(n)}&=\frac{k_i^+(\mathbf{A^*})}{\sum_{j(\neq i)}\frac{z_j^{(n-1)}}{1+x_i^{(n-1)}z_j^{(n-1)}+y_i^{(n-1)}w_j^{(n-1)}}}\quad\forall\:i,\\
y_i^{(n)}&=\frac{k_i^-(\mathbf{A^*})}{\sum_{j(\neq i)}\frac{w_j^{(n-1)}}{1+x_i^{(n-1)}z_j^{(n-1)}+y_i^{(n-1)}w_j^{(n-1)}}}\quad\forall\:i,\\
z_i^{(n)}&=\frac{h_i^+(\mathbf{A^*})}{\sum_{j(\neq i)}\frac{x_j^{(n-1)}}{1+x_j^{(n-1)}z_i^{(n-1)}+y_j^{(n-1)}w_i^{(n-1)}}}\quad\forall\:i,\\
w_i^{(n)}&=\frac{h_i^-(\mathbf{A^*})}{\sum_{j(\neq i)}\frac{y_j^{(n-1)}}{1+x_j^{(n-1)}z_i^{(n-1)}+y_j^{(n-1)}w_i^{(n-1)}}}\quad\forall\:i.
\end{align}

In order for the iterative recipe to converge, for each dataset we need to choose an appropriate initial guess.

\clearpage

\subsection{Resolution of the SDCM-FT}

In the case of the SDCM-FT, the iterative recipe that has to be solved can be written as

\begin{align}
x_i^{(n)}&=\frac{k_i^+(\mathbf{A^*})}{\sum_{j(\neq i)}|a_{ij}^*|\frac{z_j^{(n-1)}}{x_i^{(n-1)}z_j^{(n-1)}+y_i^{(n-1)}w_j^{(n-1)}}}\quad\forall\:i,\\
y_i^{(n)}&=\frac{k_i^-(\mathbf{A^*})}{\sum_{j(\neq i)}|a_{ij}^*|\frac{w_j^{(n-1)}}{x_i^{(n-1)}z_j^{(n-1)}+y_i^{(n-1)}w_j^{(n-1)}}}\quad\forall\:i,\\
z_i^{(n)}&=\frac{h_i^+(\mathbf{A^*})}{\sum_{j(\neq i)}|a_{ij}^*|\frac{x_j^{(n-1)}}{x_j^{(n-1)}z_i^{(n-1)}+y_j^{(n-1)}w_i^{(n-1)}}}\quad\forall\:i,\\
w_i^{(n)}&=\frac{h_i^-(\mathbf{A^*})}{\sum_{j(\neq i)}|a_{ij}^*|\frac{y_j^{(n-1)}}{x_j^{(n-1)}z_i^{(n-1)}+y_j^{(n-1)}w_i^{(n-1)}}}\quad\forall\:i.
\end{align}

Table~\ref{tab:MMOGfixedpoint} shows the time employed by our iterative algorithm to converge as well as the accuracy in estimating the constraints defining the SDCM and the SDCM-FT, on each snapshot of the MMOG dataset.\\

\begin{table}[t!]
\centering
\begin{tabular}{c||c|c|c||c|c|c}
\hline
\multicolumn{1}{c||}{} & \multicolumn{3}{c||}{\textbf{SDCM}} & \multicolumn{3}{c}{\textbf{SDCM-FT}} \\ 
\hline
\hline
\textit{MMOG} & MAE & MRE & Time (s) & MAE & MRE & Time (s) \\
\hline
\hline
\textit{Day 10} & $\simeq$ 0.1057 & $\simeq$ 0.0301 & $\simeq$ 3.790 & $\simeq$ 0.0048 & $\simeq$ 0.0033 & $\simeq$ 5.283 \\
\hline
\textit{Day 11} & $\simeq$  0.1210 & $\simeq$ 0.0302 & $\simeq$ 4.423 & $\simeq$ 0.0078 & $\simeq$ 0.0047 & $\simeq$ 5.885 \\
\hline
\textit{Day 12} & $\simeq$ 0.1191 & $\simeq$ 0.0312 & $\simeq$ 4.546 & $\simeq$ 0.0314 & $\simeq$ 0.0141 & $\simeq$ 6.279 \\
\hline
\textit{Day 13} & $\simeq$ 0.1248 & $\simeq$ 0.0304 & $\simeq$ 5.949 & $\simeq$ 0.0389 & $\simeq$ 0.0138 & $\simeq$ 6.553 \\
\hline
\textit{Day 14} & $\simeq$ 0.1396 & $\simeq$ 0.0282 & $\simeq$ 8.581 & $\simeq$ 0.0284 & $\simeq$ 0.0124 & $\simeq$ 7.175 \\
\hline
\textit{Day 15} & $\simeq$ 0.1399 & $\simeq$ 0.0287 & $\simeq$ 9.502 & $\simeq$ 0.0396 & $\simeq$ 0.0097 & $\simeq$ 7.646 \\
\hline
\textit{Day 16} & $\simeq$ 0.1435 & $\simeq$ 0.0348 & $\simeq$ 9.879 & $\simeq$ 0.0253 & $\simeq$ 0.0162 & $\simeq$ 9.925 \\
\hline
\textit{Day 17} & $\simeq$ 0.1372 & $\simeq$ 0.0305 & $\simeq$ 10.77 & $\simeq$ 0.0348 & $\simeq$ 0.0179 & $\simeq$ 9.663 \\
\hline
\textit{Day 18} & $\simeq$  0.1332 & $\simeq$ 0.0324 & $\simeq$ 11.46 & $\simeq$ 0.0306 & $\simeq$ 0.0180 & $\simeq$ 9.494 \\
\hline
\textit{Day 19} & $\simeq$ 0.1575 & $\simeq$ 0.0281 & $\simeq$ 11.92 & $\simeq$ 0.0298 & $\simeq$ 0.0230 & $\simeq$ 11.94 \\
\hline
\textit{Day 20} & $\simeq$ 0.1313 & $\simeq$ 0.0332 & $\simeq$ 12.49 & $\simeq$ 0.0257 & $\simeq$ 0.0219 & $\simeq$ 10.30 \\
\hline
\end{tabular}
\caption{Performance of our iterative recipe to solve the systems of equations defining the SDCM and the SDCM-FT, for each snapshot of the MMOG dataset. The table provides only the results corresponding to the best choice of the initial conditions.}
\label{tab:MMOGfixedpoint}
\end{table}

\clearpage

\hypertarget{AppD}{}
\section{Appendix D.\\ Sampling ensembles}

As each of our benchmarks treats edges independently, the ensemble it induces can be sampled quite straightforwardly as follows (only the algorithms for the SDCM and the SDCM-FT are shown).

\begin{algorithm}[h!]
\caption{Pseudocode for sampling the ensemble induced by the SDCM}
\begin{algorithmic}
\item[\hspace{1.4pt} 1:] \textbf{A=0};
\item[\hspace{1.4pt} 2:] \textbf{for} $i=1\dots N$ \textbf{do}
\item[\hspace{1.4pt} 3:] \hspace{15pt}\textbf{for} $j=1\dots N$ \textbf{do}
\item[\hspace{1.4pt} 4:] \hspace{15pt}$u=\text{RandomUniform}[0,1]$;
\item[\hspace{1.4pt} 5:] \hspace{30pt}\textbf{if} $u\le p_{ij}^-$ \textbf{then}
\item[\hspace{1.4pt} 6:] \hspace{45pt}$a_{ij}=-1$;
\item[\hspace{1.4pt} 7:] \hspace{30pt}\textbf{else if} $p_{ij}^-<u\le p_{ij}^-+p_{ij}^+$ \textbf{then}
\item[\hspace{1.4pt} 8:] \hspace{45pt}$a_{ij}=+1$;
\item[\hspace{1.4pt} 9:] \hspace{30pt}\textbf{end}
\item[10:] \hspace{15pt}\textbf{end}
\item[11:] \textbf{end}
\end{algorithmic} 
\label{alg:SDCM}
\end{algorithm}

\begin{algorithm}[h!]
\caption{Pseudocode for sampling the ensemble induced by the SDCM-FT}
\begin{algorithmic}
\item[\hspace{1.4pt} 1:] \textbf{A} $\leftarrow$ $N\times N$ matrix with $0,1$ entries;
\item[\hspace{1.4pt} 2:] \textbf{for} $i=1\dots N$ \textbf{do}
\item[\hspace{1.4pt} 3:] \hspace{15pt}\textbf{for} $j=1\dots N$ \textbf{do}
\item[\hspace{1.4pt} 4:] \hspace{30pt}\textbf{if} $a_{ij}=1$ \textbf{then}
\item[\hspace{1.4pt} 5:] \hspace{45pt}$u=\text{RandomUniform}[0,1]$;
\item[\hspace{1.4pt} 6:] \hspace{45pt}\textbf{if} $u\le p_{ij}^-$ \textbf{then}
\item[\hspace{1.4pt} 7:] \hspace{60pt}$a_{ij}=-1$;
\item[\hspace{1.4pt} 8:] \hspace{45pt}\textbf{otherwise}
\item[\hspace{1.4pt} 9:] \hspace{60pt}$a_{ij}=+1$;
\item[10:] \hspace{45pt}\textbf{end}
\item[11:] \hspace{30pt}\textbf{end}
\item[12:] \hspace{15pt}\textbf{end}
\item[13:] \textbf{end}
\end{algorithmic} 
\label{alg:SDCMFT}
\end{algorithm}

\clearpage

\hypertarget{AppE}{}
\section{Appendix E.\\Inspecting dyadic motifs on binary, directed, signed networks}\label{AppE}

This Appendix is devoted to calculate the expected values and the variances of the abundances of the single dyads, i.e.

\begin{align}
L^\rightarrow_+&=\sum_{i=1}^N\sum_{j(>i)}[a_{ij}^+(1-a_{ji}^+-a_{ji}^-)+a_{ji}^+(1-a_{ij}^+-a_{ij}^-)]
\end{align}
and

\begin{align}
L^\rightarrow_-&=\sum_{i=1}^N\sum_{j(>i)}[a_{ij}^-(1-a_{ji}^+-a_{ji}^-)+a_{ji}^-(1-a_{ij}^+-a_{ij}^-)].
\end{align}

The expected abundance of the positive, single dyad reads

\begin{align}
\langle L^\rightarrow_+\rangle&=\sum_{i=1}^N\sum_{j(>i)}[p_{ij}^+(1-p_{ji}^+-p_{ji}^-)+p_{ji}^+(1-p_{ij}^+-p_{ij}^-)]
\end{align}
while its variance reads

\begin{align}
\text{Var}[L^\rightarrow_+]&=\sum_{i=1}^N\sum_{j(>i)}\text{Var}[a_{ij}^+(1-a_{ji}^+-a_{ji}^-)+a_{ji}^+(1-a_{ij}^+-a_{ij}^-)]\nonumber\\
&=\sum_{i=1}^N\sum_{j(>i)}\text{Var}[a_{ij}^+(1-a_{ji}^+-a_{ji}^-)]+\text{Var}[a_{ji}^+(1-a_{ij}^+-a_{ij}^-)]+2\text{Cov}[a_{ij}^+(1-a_{ji}^+-a_{ji}^-),a_{ji}^+(1-a_{ij}^+-a_{ij}^-]\nonumber\\
&=\sum_{i=1}^N\sum_{j(>i)}[p_{ij}^+q_{ji}(1-p_{ij}^+q_{ji})+p^+_{ji}q_{ij}(1-p^+_{ji}q_{ij})-2p_{ij}^+q_{ji}p_{ji}^+q_{ij}]\nonumber\\
&=\sum_{i=1}^N\sum_{j(>i)}[p_{ij}^+q_{ji}(1-p_{ij}^+q_{ji}-p_{ji}^+q_{ij})+q_{ij}p^+_{ji}(1-p^+_{ij}q_{ji}-p_{ji}^+q_{ij})]
\end{align}
with $q_{ji}=1-p_{ij}^+(1-p_{ji}^+-p_{ji}^-)$ and $q_{ij}=1-p_{ji}^+(1-p_{ij}^+-p_{ij}^-)$, since

\begin{align}
\text{Cov}[a_{ij}^+(1-a_{ji}^+-a_{ji}^-),a_{ji}^+(1-a_{ij}^+-a_{ij}^-)]&=\langle a_{ij}^+(1-a_{ji}^+-a_{ji}^-)a_{ji}^+(1-a_{ij}^+-a_{ij}^-)\rangle-\langle a_{ij}^+(1-a_{ji}^+-a_{ji}^-)\rangle\langle a_{ji}^+(1-a_{ji}^+-a_{ji}^-)\rangle\nonumber\\
    &=-p_{ij}^+(1-p_{ji}^+-p_{ji}^-)p_{ij}^+(1-p_{ij}^+-p_{ij}^-)
\end{align}
as the events $a_{ij}^+(1-a_{ji}^+-a_{ji}^-)=1$ and $a_{ji}^+(1-a_{ij}^+-a_{ij}^-)=1$ are mutually exclusive.\\

Similarly, the expected abundance of the negative, single dyad reads

\begin{align}
\langle L^\rightarrow_-\rangle&=\sum_{i=1}^N\sum_{j>i}[p_{ij}^-(1-p_{ji}^+-p_{ji}^-)+p_{ji}^-(1-p_{ij}^+-p_{ij}^-)]
\end{align}
while its variance reads

\begin{align}
\text{Var}[L^\rightarrow_-]&=\sum_{i=1}^N\sum_{j>i}\text{Var}[a_{ij}^-(1-a_{ji}^+-a_{ji}^-)+a_{ji}^-(1-a_{ij}^+-a_{ij}^-)]\nonumber\\
&= \sum_{i=1}^N\sum_{j>i}\text{Var}[a_{ij}^-(1-a_{ji}^+-a_{ji}^-)]+\text{Var}[a_{ji}^-(1-a_{ij}^+-a_{ij}^-)]+2\text{Cov}[a_{ij}^-(1-a_{ji}^+-a_{ji}^-),a_{ji}^-(1-a_{ij}^+-a_{ij}^-)]\nonumber\\
&=\sum_{i=1}^N\sum_{j>i}[p_{ij}^-q_{ji}(1-p_{ij}^-q_{ji})+p_{ji}^-q_{ij}(1-q_{ij}p^-_{ji})-2p_{ij}^-q_{ji}p_{ji}^-q_{ij}]\nonumber\\
&= \sum_{i=1}^N\sum_{j>i}[p_{ij}^-q_{ji}(1-p_{ij}^-q_{ji}-p_{ji}^-q_{ij})+q_{ij}p^-_{ji}(1-p^+_{ij}q_{ji}-p_{ji}^-q_{ij})].
\end{align}
with $q_{ji}=1-p_{ij}^-(1-p_{ji}^+-p_{ji}^-)$ and $q_{ij}=1-p_{ji}^-(1-p_{ij}^+-p_{ij}^-)$, since

\begin{align}
\text{Cov}[a_{ij}^-(1-a_{ji}^+-a_{ji}^-),a_{ji}^-(1-a_{ij}^+-a_{ij}^-)] &= \langle a_{ij}^-(1-a_{ji}^+-a_{ji}^-)a_{ji}^-(1-a_{ij}^+-a_{ij}^-)\rangle-\langle a_{ij}^-(1-a_{ji}^+-a_{ji}^-)\rangle\langle a_{ji}^-(1-a_{ji}^+-a_{ji}^-)\rangle\nonumber\\
&=-p_{ij}^-(1-p_{ji}^+-p_{ji}^-)p_{ij}^-(1-p_{ij}^+-p_{ij}^-)
\end{align}
as the events $a_{ij}^-(1-a_{ji}^+-a_{ji}^-)=1$ and $a_{ji}^-(1-a_{ij}^+-a_{ij}^-)=1$ are mutually exclusive.\\

\begin{figure}[t!]
\centering
\includegraphics[width=1\textwidth]{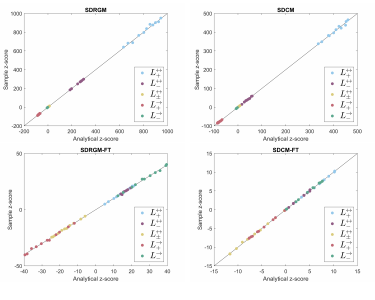}
\caption{Analytical VS sample $z$-scores of both the reciprocal and single dyads, for each snapshot of the MMOG dataset. Each sample is defined by 1.000 realisations of the corresponding benchmark.}
\label{fig:9}
\end{figure}

As Figure~\ref{fig:9} shows, the analytical calculations match their sample counterparts.

\end{document}